\documentclass{article}

\usepackage{arxiv}

\renewcommand{\shorttitle}{Jurisdiction-Siloed Disclosure in Privacy Policies}

\usepackage[utf8]{inputenc}
\usepackage[T1]{fontenc}
\usepackage{hyperref}
\usepackage{url}
\usepackage{microtype}

\usepackage{array}
\usepackage{placeins}
\usepackage{booktabs}
\usepackage{graphicx}

\usepackage{amsmath}
\usepackage{amssymb}

\usepackage{pifont}

\usepackage{tikz}
\usetikzlibrary{patterns, positioning, arrows.meta, calc, fit, decorations.pathreplacing, matrix}

\usepackage{natbib}
\newcommand{\cmark}{\ding{51}}

\begin{document}

\title{Jurisdiction as Structural Barrier: How Privacy Policy Organization May Reduce Visibility of Substantive Disclosures}

\author{
Thomas Brackin \\
Independent Researcher \\
Seattle, WA, USA \\
\texttt{tebrackin@outlook.com}
}

\date{January 2026}

\begin{abstract}
Privacy policies are supposed to provide notice. But what if substantive information appears only where users skip it? We identify a structural pattern we call \emph{jurisdiction-siloed disclosure}: information about data practices appearing in specific, actionable form only within regional compliance sections labeled ``California Residents'' or ``EU/UK Users,'' while general sections use vague or qualified language for the same practices.

Our audit of 123 major companies identifies 282 potential instances across 77 companies (62.6\% of this purposive sample). A conservative estimate restricted to practice categories validated against OPP-115 human annotations finds 138 instances across 54 companies (44\%); post-2018 categories central to our findings await independent validation. If users skip jurisdiction-labeled sections as information foraging theory predicts, users outside regulated jurisdictions would receive less specific information about practices affecting them---a transparency failure operating through document architecture rather than omission.

We propose \emph{universal substantive disclosure}: practices affecting all users should appear in the main policy body, with regional sections containing only procedural rights information. This standard finds support in analogous disclosure regimes (securities, truth-in-lending, nutritional labeling) where material information must reach all affected parties. Regulators could operationalize this through the FTC's ``clear and conspicuous'' standard and GDPR transparency principles.

This work is hypothesis-generating: we establish that the structural pattern exists and ground the transparency concern in behavioral theory, but direct measurement of jurisdiction-specific section skipping remains the critical validation priority. We release our methodology and annotated dataset to enable replication.
\end{abstract}

\maketitle


\section{Introduction}

Privacy policies are supposed to provide notice. The ``notice and choice'' framework assumes users can read, understand, and act upon disclosed data practices. Scholars have thoroughly documented why this assumption fails for individual users: policies are too long~\cite{mcdonald2008cost}, written at inaccessible reading levels, and impose unrealistic cognitive burdens~\cite{solove2013consent}. Only 9\% of users report always reading policies before accepting them~\cite{pew2019privacy}.

But even users who read policies cover-to-cover may not receive complete notice. We identify a candidate structural pattern that may undermine transparency through organization rather than omission: \emph{jurisdiction-siloed disclosure}, where substantive information about data practices appears in specific, structured form only within sections targeting users in specific legal jurisdictions, while general sections provide at most qualified or conditional language for the same practices.

\subsection{The Problem}

Consider a privacy policy with a section titled ``Your California Privacy Rights.'' A reasonable user in Ohio, seeing this heading, might skip it---the section appears to contain information relevant only to Californians. But what if that section contains the only explicit disclosure that the company sells user data to advertisers? If the Ohio user engages in the jurisdiction-based filtering that information foraging theory predicts, they would be unlikely to receive notice of this practice through normal policy navigation.

Our argument rests on an informed inference from established behavioral research: information foraging theory~\cite{pirolli1999information}, heading-based navigation studies~\cite{vu2007users}, and documented selective reading patterns~\cite{obar2020biggest} collectively predict that users are likely to skip sections whose headings signal geographic irrelevance. While direct measurement of jurisdiction-specific section skipping remains valuable future work (Section~\ref{sec:validation}), the inference is well-grounded in convergent evidence from multiple research traditions.

This is not a hypothetical. Our audit of 123 major companies suggests that a majority may exhibit this pattern, with substantive disclosures (data sales, biometric collection, automated profiling) appearing in specific form only within jurisdiction-specific sections, while general sections use less precise language.

Current law permits this structure. CCPA requires California-specific disclosures; GDPR mandates EU-specific information; BIPA requires Illinois-specific notices about biometric data. A company can achieve technical compliance by placing required disclosures in jurisdiction-specific sections. But if users skip jurisdiction-labeled sections as information foraging theory predicts, the effect would be that users outside those jurisdictions, who lack legal rights but may be subject to identical practices, would receive systematically incomplete information.

The architecture may satisfy the letter of jurisdiction-specific laws while limiting the informational value for users outside specified jurisdictions. A user in Texas whose facial geometry is captured and sold may be unlikely to encounter specific notice of this practice under typical navigation patterns because the detailed disclosure appears only in an Illinois BIPA compliance section, with at most circumlocutory language in the general policy. A user in Florida whose data is sold to advertisers may be unlikely to encounter this information through standard policy browsing because the explicit sale disclosure appears only under ``California Consumer Privacy Act,'' while general sections may use imprecise language that does not clearly communicate data sales. The accident of geography determines not what rights one can exercise---that is the legitimate function of jurisdiction-specific law---but what information one receives about practices affecting them.

\subsection{Universal Substantive Disclosure}

We propose a normative standard grounded in a distinction between substantive and procedural content (developed in Section~\ref{sec:proposal}): substantive practices affecting all users should be disclosed universally, while regional sections should contain only procedural information. The disclosure that data \emph{is} sold should appear in the main body of the policy; the California section provides the opt-out mechanism.

This principle does not require that all users have identical rights. Jurisdiction-specific protections remain jurisdiction-specific. It requires only that all users have access to the same information about what companies do with their data. The asymmetry in legal protection should not be compounded by an asymmetry in disclosure.

\subsection{Contributions}

\begin{enumerate}
    \item Identification and operationalization of \emph{jurisdiction-siloed disclosure} as a candidate structural pattern warranting systematic investigation, with preliminary evidence suggesting prevalence among major companies.
    \item \emph{Universal substantive disclosure} as a proposed normative standard for privacy policy organization, grounded in philosophical foundations of autonomy, fairness, and truthfulness.
    \item A substantive/procedural framework and annotation methodology enabling independent researchers to test for jurisdiction-siloed disclosure in their own samples.
    \item Case studies illustrating how the pattern operates in practice, with varying levels of external verification.
    \item An annotated dataset of 123 major company privacy policies released to enable independent replication, validation, and extension.
\end{enumerate}

This work represents a \emph{hypothesis-generating} contribution rather than a definitive prevalence study. We release our methodology and data to invite independent validation of both the phenomenon's existence and its frequency.

Section~\ref{sec:validation} details the validation studies needed to confirm these findings.

\section{Background and Related Work}

\subsection{The Notice-and-Choice Framework}

Privacy regulation in the United States relies on ``notice and choice'': companies disclose data practices; users decide whether to accept. This framework assumes meaningful notice: that users receive, read, and understand disclosures.

Decades of research document this assumption's failure. McDonald and Cranor~\cite{mcdonald2008cost} estimated that reading the privacy policy for every website an average user visits would require approximately 244 hours per year. Solove~\cite{solove2013consent} articulated the ``consent dilemma'': meaningful consent requires comprehension that the notice-and-choice framework structurally cannot deliver. Reidenberg et al.~\cite{reidenberg2015disagreeable} demonstrated that even privacy experts disagree on the interpretation of policy language, suggesting that textual ambiguity undermines notice independently of user effort. Cranor~\cite{cranor2012necessary} argued that standardized notice mechanisms are necessary but not sufficient for privacy protection, a thesis our structural findings reinforce. In practice, policies function primarily as instruments of legal compliance rather than tools for informed decision-making.

\subsection{Beyond Comprehension: The Organization Problem}

Prior work has focused primarily on \emph{comprehension} barriers: policies are too long, too complex, written at too high a reading level. Our contribution identifies a distinct barrier: \emph{organizational} choices that create conditions where disclosures may be difficult to find for users employing the selective reading strategies documented in prior research.

This problem is structural rather than linguistic. A disclosure written in plain English, at a sixth-grade reading level, with perfect clarity, may still fail to provide notice if it appears only in a section users are likely to skip. Empirical research documents that 74\% of users skip privacy policies entirely, and among those who engage, average reading time is under 90 seconds for documents requiring 29--32 minutes~\cite{obar2020biggest}. Users navigate selectively, examining headings to determine relevance before engaging with content~\cite{vu2007users}. This suggests section headings function as gatekeepers, causing users to skip sections whose headings signal irrelevance. Eye-tracking research confirms this pattern: Steinfeld found that users who click to view policies typically skim rather than read, and most skip the policy entirely when given the option~\cite{steinfeld2016agree}.

Information foraging theory~\cite{pirolli1999information} explains this behavior: users rely on ``information scent''---environmental cues that guide navigation toward relevant content. When scent is weak or absent, users abandon content rather than exploring exhaustively. Section headings like ``Your California Privacy Rights'' provide strong \emph{negative} information scent for users outside those jurisdictions, triggering the satisficing behavior~\cite{simon1956rational} that causes users to skip content perceived as irrelevant.

McDonald et al.~\cite{mcdonald2009comparative} found that layered notices can effectively hide information and reduce transparency when users cannot find needed information in summary layers. Schaub et al.~\cite{schaub2015designspace} developed a design space framework for privacy notices identifying timing, channel, modality, and control as critical dimensions affecting notice effectiveness. Their emphasis on information presentation beyond content (where and when disclosures appear) provides theoretical grounding for structural transparency concerns. A disclosure that appears at the ``wrong time'' (in a section the user has decided to skip based on heading cues) or through a misaligned ``channel'' (a section framed as geographically irrelevant) fails the design space criteria for effective notice regardless of content quality. We hypothesize an analogous phenomenon: jurisdiction-specific sections may function as visibility barriers when substantive disclosures appear only within them. These sections satisfy the conditions established in the literature for content users systematically overlook; our contribution is to document the structural phenomenon that creates this risk, with behavioral validation requirements detailed in Section~\ref{sec:behavioral_validation}.

\subsection{Dark Patterns and Structural Transparency}

Gray et al.~\cite{gray2018dark} formalized a taxonomy of dark patterns, design choices that subvert user autonomy, identifying five strategies including \emph{sneaking} (concealing relevant information) and \emph{interface interference} (manipulated UI hierarchy). The FTC's 2022 staff report~\cite{ftc2022darkpatterns} brought regulatory attention to these practices, cataloguing dark patterns including design elements that obscure or delay material disclosures. Research has documented dark patterns in consent interfaces~\cite{nouwens2020dark} and cookie banners~\cite{machuletz2020cookie}, and significant usability barriers in data deletion and opt-out processes~\cite{habib2019empirical}.

We identify a phenomenon we term \emph{structural visibility reduction}: jurisdiction-siloed disclosure operates through document architecture rather than interface manipulation, placing it outside traditional dark patterns research while potentially producing analogous effects. Of Gray et al.'s categories, ``sneaking,'' defined as ``an attempt to hide, disguise, or delay the divulging of information that has relevance to the user,'' is conceptually closest, though jurisdiction-siloed disclosure involves \emph{organizing} rather than \emph{omitting} information~\cite{gray2018dark}.

The dark patterns literature debates whether intentionality is necessary for classification~\cite{mathur2019dark,luguri2021shining}. Jurisdiction-siloed disclosure likely emerges from compliance motivations rather than deliberate manipulation. But if dark patterns are defined by \emph{effects} rather than \emph{intent}, structural patterns may warrant consideration. We propose that jurisdiction-siloed disclosure represents a \emph{structural analog} to interface dark patterns, operating at the level of policy organization rather than UI design.

\subsection{Regulatory Context}

Privacy regulations globally specify \emph{what} companies must disclose but not \emph{where} those disclosures must appear within a policy's organizational structure. This content-without-structure approach creates a regulatory gap: companies can technically comply with disclosure mandates while placing substantive information in sections that users are likely to skip based on geographic framing.

\subsubsection{CCPA: Content Without Structure}

The California Consumer Privacy Act requires disclosure of specific categories: personal information collected, sold, and shared; purposes; and consumer rights~\cite{ccpa2018,cppa2023regs}. Critically, the statute does not mandate that this information appear in a separate California section; separate sections are an \emph{industry practice}, not a statutory requirement. A company could satisfy CCPA by including required disclosures in its main policy, visible to all users, while placing only California-specific \emph{procedures} (opt-out mechanisms, authorized agent processes) in a dedicated section. The choice to place \emph{substantive} disclosures exclusively in California sections reflects organizational decisions, not legal necessity.

\subsubsection{GDPR Layered Notice}

GDPR's transparency guidelines (WP260) recommend layered notices with first-layer content including ``information on the processing which has the most impact on the data subject and processing which could surprise them''~\cite{wp260}. The ``easily accessible'' requirement of GDPR Article 12, interpreted as meaning users ``should not have to seek out the information,'' raises the question of whether placing substantive disclosures in sections framed as geographically irrelevant creates precisely the search burden that transparency principles aim to prevent. WP260 does not explicitly address jurisdiction-siloed placement, but the guidance suggests material practices should appear where users are likely to find them.

\subsubsection{Global Pattern}

This content-without-structure gap extends globally. Brazil's LGPD (2018) requires ``clear, accurate, and easily accessible information''~\cite{lgpd2018}; China's PIPL (2021) mandates disclosure ``in an eye-catching manner''~\cite{pipl2021}. Both specify disclosure \emph{content} without mandating organizational placement. China's 2025 Network Data Security Management Regulations introduce a ``dual-list'' format requirement (an emerging exception that addresses disclosure \emph{structure}), though even this does not prohibit jurisdiction-siloed placement.

\subsubsection{The Regulatory Gap}

To our knowledge, no regulator has directly addressed whether organizational choices (placing substantive disclosures in sections users are likely to skip based on geographic framing) violate transparency requirements. Current enforcement focuses on disclosure \emph{content} (whether required information appears somewhere) rather than disclosure \emph{structure} (whether that information appears where users are likely to locate it). Our contribution is to identify jurisdiction-siloed disclosure as a structural transparency problem that existing frameworks do not explicitly address, though existing principles suggest it warrants regulatory attention.

\subsection{Computational Privacy Policy Analysis}

A parallel research program has developed automated tools for privacy policy analysis. The OPP-115 corpus~\cite{wilson2016creation} established foundational annotation methods, using three independent law student annotators to classify 115 policies into fine-grained data practice categories. Subsequent work introduced Polisis for automated segment classification~\cite{harkous2018polisis}, PolicyLint for contradiction detection~\cite{andow2019policylint}, and the PrivaSeer corpus enabling million-policy-scale analysis~\cite{srinath2021privaseer}. Zimmeck et al.~\cite{zimmeck2019maps} extended compliance analysis to code-policy discrepancies at scale, demonstrating that automated methods can detect mismatches between stated privacy practices and actual app behavior across a million applications.

These tools focus on \emph{content} extraction: identifying what practices policies disclose. Our contribution addresses \emph{structural} transparency: whether disclosures appear where users will find them. Existing tools detect that a policy mentions data sales; they do not assess whether that disclosure appears universally or within jurisdiction-specific content users may skip. Structural analysis is required.

\section{Universal Substantive Disclosure}
\label{sec:proposal}

We propose that privacy policies should satisfy a structural requirement we call \emph{universal substantive disclosure}: substantive practices affecting all users should be disclosed to all users. This section develops the philosophical, legal, and practical foundations for this standard.

\subsection{The Substantive/Procedural Distinction}
\label{sec:distinction}

We distinguish two types of content in jurisdiction-specific sections: procedural and substantive.

\emph{Procedural disclosures} describe how users in specific jurisdictions can exercise rights. These include opt-out links, data access request forms, supervisory authority contacts, consent withdrawal mechanisms, and authorized agent procedures. Procedural content is appropriately jurisdiction-specific because the procedures differ by legal regime. It makes no sense to direct a Texas resident to file a complaint with the Irish Data Protection Commission.

\emph{Substantive disclosures} describe what the company does: whether they sell data, collect biometrics, use automated profiling. These practices typically affect all users regardless of jurisdiction. If a company sells user data to data brokers, this practice affects users in Texas and California alike.

Universal substantive disclosure requires that substantive content appear universally while permitting procedural content to remain jurisdiction-specific.

\subsection{Philosophical Foundations}

The case for universal disclosure rests on three foundational principles: autonomy, fairness, and truthfulness.

First, the notice-and-choice framework presupposes that individuals can make meaningful decisions about their data based on disclosed information. This presupposition of autonomy applies regardless of jurisdiction. A Texas resident deciding whether to use a service that sells biometric data has the same interest in knowing this as an Illinois resident---the practice affects both identically. To condition disclosure on the availability of legal remedy is to abandon the transparency rationale for privacy policies entirely.

Second, selective disclosure creates information asymmetries based solely on residence, raising concerns of fairness. Jiang et al.'s ``Principle of Minimum Asymmetry'' holds that privacy-preserving systems should minimize imbalances between data owners and data collectors regarding knowledge about practices and personal data~\cite{jiang2002approximate}. Jurisdiction-siloed disclosure violates this principle: users in jurisdictions with strong privacy laws receive comprehensive notice; users elsewhere receive systematically incomplete information. This asymmetry correlates with existing inequalities: states with stronger privacy laws tend to be wealthier and more politically organized. The result is a disclosure regime that provides more information to those who least need protection while concealing practices from those with the fewest resources to discover them independently.

Third, privacy policies purport to describe company practices, invoking an obligation of truthfulness that extends beyond mere accuracy to informational completeness. Grice's Cooperative Principle~\cite{grice1975logic} holds that communicative exchanges presuppose adherence to conversational maxims, including the maxim of \emph{quantity}: ``Make your contribution as informative as is required.'' When a speaker appears to violate this maxim, listeners draw \emph{conversational implicatures}, inferences about what the speaker must have meant by their apparent omission~\cite{levinson1983pragmatics}. Crucially, a speaker who provides partial information creates an implicature that no additional relevant information exists; silence about a topic implies that topic does not apply.

Privacy policies function as communicative acts subject to these pragmatic norms. A user reading a privacy policy reasonably assumes, consistent with the maxim of quantity, that the policy discloses material practices affecting them. A policy that describes biometric collection in specific terms only in an Illinois section creates a conversational implicature for non-Illinois readers: the absence of clear biometric disclosure in the main body implies their biometrics are not collected. The user who reads a privacy policy from start to finish and finds no specific mention of biometric collection is entitled to conclude that no such collection occurs, even if oblique references to ``information derived from facial appearance'' exist---such circumlocutory language does not convey the same meaning as explicit biometric disclosure. This is not a logical entailment but a \emph{pragmatic} inference licensed by the communicative context, exactly the inference Gricean theory predicts.

Philosophers distinguish between \emph{lying} (asserting what one believes to be false) and \emph{misleading} (creating false impressions through technically true statements or strategic omissions)~\cite{saul2012lying}. Jurisdiction-siloed disclosure falls squarely within the misleading category: every individual statement may be accurate, yet the policy's organizational structure creates false impressions about what practices affect which users. This is deception through architecture rather than assertion---and, we argue, no less a violation of truthfulness norms for being structural rather than propositional.

\subsection{Legal Analogies}

Universal substantive disclosure finds support in analogous disclosure regimes. Securities law requires that material information be disclosed to all investors, not merely those in jurisdictions with stronger enforcement; the rationale is that markets function efficiently only when all participants have access to the same material facts. Truth-in-lending regulations mandate uniform disclosure of credit terms regardless of state usury laws; a lender cannot disclose an APR only in states where that rate would be illegal, because the disclosure obligation is independent of the substantive regulation. Similarly, nutritional labeling requires uniform ingredient disclosure regardless of whether a particular ingredient is restricted in the consumer's jurisdiction; a food manufacturer cannot omit trans fat content from labels in states without trans fat bans.

These analogies suggest a general principle: when disclosure regimes are designed to enable informed decision-making, the disclosure obligation should not be conditioned on the existence of substantive rights.

\subsection{The Proposed Standard}

Under universal substantive disclosure, a privacy policy would satisfy the following requirements:

\begin{enumerate}
    \item \emph{Universal disclosure of practices.} All substantive data practices (collection, use, sharing, sale, automated processing, sensitive data handling) appear in the main body, visible to all users regardless of jurisdiction.
    \item \emph{Regional sections contain only procedures.} Jurisdiction-specific sections describe how residents can exercise rights under applicable law: opt-out mechanisms, access request procedures, supervisory authority contacts.
    \item \emph{No substantive-only regional content.} A disclosure that appears only in a regional section is presumptively non-compliant with universal substantive disclosure.
    \item \emph{Cross-references permitted.} Regional sections may reference substantive disclosures that appear elsewhere (``As described in Section 3, we sell personal information. California residents can opt out by...'').
\end{enumerate}

\subsection{Defining Universal Practices}

The proposed standard requires that ``substantive practices affecting all users'' be disclosed universally. But what constitutes a practice that ``affects all users''? This section provides an operational definition and guidance for distinguishing universal practices from genuinely jurisdiction-specific ones.

\subsubsection{Operational Definition}

A data practice should be presumed to affect all users, and thus require universal disclosure, unless the company can demonstrate otherwise. Specifically, a practice ``affects all users'' if:

\begin{enumerate}
    \item[(a)] Technical infrastructure does not support jurisdiction-specific variants. Most data processing systems operate uniformly across users. A facial recognition database either contains a user's image or it does not; a data broker sale either includes a user's information or it does not. If a company's infrastructure processes all users identically, the practice affects all users.

    \item[(b)] The policy lacks an explicit statement limiting the practice to certain jurisdictions. When a disclosure appears only in a California section without language such as ``This practice applies only to California residents'' or ``We do not sell data from users outside California,'' the reasonable inference is that the practice is universal but the disclosure is not.

    \item[(c)] Regulatory variation does not explain selective disclosure. Some practices are disclosed regionally because regulations \emph{require} disclosure in specific jurisdictions. But if the underlying practice is not itself jurisdiction-specific, meaning the company would engage in the same practice regardless of where the user resides, then the regulatory trigger for disclosure should not limit where the disclosure appears.
\end{enumerate}

This framework creates a rebuttable presumption of universality. Companies that genuinely limit practices to specific jurisdictions can overcome the presumption by explicitly stating the geographic limitation in their policies.

\subsubsection{When Jurisdiction-Specific Disclosure Is Appropriate}

Not all jurisdiction-specific disclosures represent concealment. Some practices genuinely vary by jurisdiction for legitimate operational, legal, or technical reasons.

\emph{Example 4: International Data Transfers (Appropriate Regional Disclosure).}

\textit{EU/UK section}: ``Personal data of EU/UK users is transferred to the United States under the EU-U.S. Data Privacy Framework. Your data is stored in our EU data centers located in Ireland and Germany.''

\textit{Main body}: ``We operate data centers in multiple regions. Your data may be processed in the country where you access our services or in other countries where we or our service providers operate.''

\textit{Analysis}: This is \emph{not} jurisdiction-siloed disclosure in the problematic sense. The EU/UK section describes a genuinely different practice: EU user data is stored in EU data centers and transferred under a specific legal mechanism (the Data Privacy Framework), while other users' data may be processed differently. The practice itself---not merely the disclosure---varies by jurisdiction.

\textit{Contrast with concealment}: If the company processed \emph{all} user data identically (storing all data in U.S. data centers, with no regional infrastructure), but disclosed this only in the EU section because GDPR requires transfer disclosure, that would be jurisdiction-siloed disclosure. The practice would be universal; only the disclosure would be regional.

\subsubsection{Guidance for Companies}

Companies seeking to comply with universal substantive disclosure should apply the following framework:

\begin{enumerate}
    \item Audit for universal practices. For each disclosure currently appearing only in a regional section, ask: Does this practice apply to users outside this jurisdiction? If yes, move the substantive disclosure to the main body.

    \item Retain regional procedural content. The mechanisms by which California residents opt out, EU residents exercise data subject rights, or Illinois residents request biometric data deletion appropriately remain in regional sections. Only the underlying practice requires universal disclosure.

    \item Explicitly state genuine limitations. If a practice genuinely applies only to certain jurisdictions, say so explicitly: ``We sell personal information only from users who access our services in California.'' Explicit limitation statements convert potentially problematic siloed disclosures into appropriately scoped regional ones.

    \item Use cross-references. Regional sections can reference universal disclosures rather than repeating them: ``As described in Section 3 above, we may sell personal information to third parties. California residents have the right to opt out of such sales by [mechanism].''
\end{enumerate}

\subsection{Addressing Objections}

We anticipate five objections to universal disclosure.

Companies might first object that universal disclosure increases compliance burden. This objection is unpersuasive. The substantive information already exists; it merely appears in the wrong section. Moving ``We sell your personal information'' from a California section to the main body requires no additional research or legal review.

A related concern involves legal complexity: jurisdiction-specific requirements might necessitate jurisdiction-specific language. This is true for procedural requirements but not for substantive disclosures. ``We collect biometric identifiers including facial geometry'' is equally accurate in every jurisdiction.

A third objection concerns information overload: if all users see all disclosures, policies become longer and more difficult to read. This concern reflects a genuine phenomenon documented in privacy policy research. McDonald and Cranor estimated that reading all applicable privacy policies would require approximately 244 hours per year~\cite{mcdonald2008cost}; Obar and Oeldorf-Hirsch found users allocate only 73 seconds on average to policies requiring 29--32 minutes for thorough reading~\cite{obar2020biggest}; follow-up research found users explicitly describe policies as overwhelming yet simultaneously irrelevant~\cite{oeldorfhirsch2019overwhelming}. Adding universal substantive disclosure would increase policy length, potentially exacerbating these barriers.

We take this concern seriously but reject it as a justification for jurisdiction-siloed disclosure for three reasons. First, the objection proves too much: if information overload justifies selective disclosure, it would equally justify omitting any practice that makes policies longer. Taken to its logical conclusion, the argument implies that shorter, less complete policies serve users better, a position inconsistent with transparency's foundational premise that users benefit from knowing what happens to their data.

Second, the problem is organizational, not volumetric. McDonald et al.\ found that layered notices can effectively hide information and reduce transparency when users cannot locate needed information in summary layers~\cite{mcdonald2009comparative}, precisely the mechanism we document. The appropriate response to complexity is better information architecture: layered disclosures with comprehensive base layers, standardized formats enabling comparison, and machine-readable policies supporting automated analysis. Universal disclosure combined with improved structure addresses both completeness and usability; jurisdiction-siloed disclosure sacrifices completeness without improving usability.

Third, the net effect on user notice favors universal disclosure. Our finding that a majority of major companies silo substantive practices (see Section~\ref{sec:prevalence}) means users outside triggering jurisdictions encounter systematically incomplete information. If universal disclosure increases policy length by 10--15\% but increases the probability that users encounter material practices from roughly 37\% to near-certainty, the tradeoff favors disclosure. Users who skim policies incompletely are more likely to encounter material disclosures through universal disclosure: a Texas user scanning Meta's policy has some probability of encountering ``We sell personal information'' if it appears in the main body, but substantially lower probability if it appears only in a collapsed California section they would likely skip based on heading cues.

We acknowledge that universal disclosure is not costless. For users in multiple regulated jurisdictions, seeing procedural rights for California, the EU, Brazil, and China in a single policy may approach cognitive overload. This argues for standardizing procedural formats (consolidated rights-exercise sections, harmonized terminology) rather than siloing substantive disclosures. The distinction matters: procedural complexity is an organizational challenge amenable to design solutions; substantive concealment is a transparency failure that design cannot remedy without disclosure.

A fourth objection notes that jurisdiction-specific sections provide genuine navigation benefits for users \emph{within} those jurisdictions. A California resident can currently turn to a single ``California Privacy Rights'' section to find consolidated information about the data categories collected, the categories sold, and the opt-out mechanisms available to them. Under universal substantive disclosure, that same resident might need to read the full policy body to locate disclosures currently consolidated in one section. We acknowledge this trade-off. However, we argue it favors universal disclosure because the substantive practices (data sales, biometric collection) affect all users, while only the procedural rights (opt-out mechanisms, deletion request processes) are genuinely jurisdiction-specific. Cross-references from regional sections to universal disclosures (``As described in Section 3, we sell personal information; California residents may opt out by...'') preserve the navigational consolidation that serves regulated users while ensuring substantive practices are visible to all.

Finally, one might argue that users in unregulated jurisdictions do not expect comprehensive disclosure and would be confused by information about practices for which they have no remedy. This paternalistic argument inverts the transparency rationale: users are denied information for their own good, because knowing might distress them without providing recourse. We reject this framing. Users are entitled to make their own judgments about services that engage in objectionable practices, regardless of legal remedy.

\begin{figure*}[t]
\centering
\resizebox{\textwidth}{!}{%
\begin{tikzpicture}[
    >=Stealth,
    section/.style={draw, rounded corners=2pt, minimum width=5.8cm, minimum height=0.7cm, align=center, font=\small},
    subsection/.style={draw, rounded corners=2pt, minimum width=5cm, minimum height=0.55cm, align=center, font=\scriptsize},
    substantive/.style={fill=black!15},
    procedural/.style={pattern=north east lines, pattern color=black!40},
    vague/.style={fill=white, draw, dashed},
    user/.style={draw, circle, minimum size=0.7cm, font=\scriptsize, thick},
    label/.style={font=\footnotesize\bfseries},
    note/.style={font=\scriptsize\itshape, text width=3.5cm, align=center},
]

\node[label] at (-4.2, 5.2) {Current Pattern};
\node[font=\scriptsize\itshape] at (-4.2, 4.8) {Jurisdiction-Siloed Disclosure};

\node[section, vague, minimum height=1.6cm] (gen-left) at (-4.2, 3.2) {};
\node[font=\footnotesize\bfseries] at (-4.2, 3.7) {General Sections};
\node[font=\scriptsize, text width=5cm, align=center] at (-4.2, 3.0) {``We may share information\\with partners\ldots''};

\node[section, minimum height=1.8cm] (ca-left) at (-4.2, 0.8) {};
\node[font=\footnotesize\bfseries] at (-4.2, 1.5) {California Privacy Rights};
\node[subsection, substantive] at (-4.2, 0.9) {Data sales disclosure};
\node[subsection, procedural] at (-4.2, 0.25) {Opt-out mechanism};

\node[section, minimum height=1.8cm] (eu-left) at (-4.2, -1.5) {};
\node[font=\footnotesize\bfseries] at (-4.2, -0.8) {EU/UK Users};
\node[subsection, substantive] at (-4.2, -1.4) {Automated profiling};
\node[subsection, procedural] at (-4.2, -2.05) {Data subject rights};

\node[section, minimum height=1.8cm] (il-left) at (-4.2, -3.7) {};
\node[font=\footnotesize\bfseries] at (-4.2, -3.0) {Illinois Residents};
\node[subsection, substantive] at (-4.2, -3.6) {Biometric collection};
\node[subsection, procedural] at (-4.2, -4.25) {Written consent};

\node[user] (user-left) at (-8.3, 1.8) {User};
\node[note, text width=2cm] at (-8.3, 0.9) {(Texas resident)};

\draw[->, thick, black] (user-left) -- node[above, font=\scriptsize] {Reads} (gen-left.west);
\draw[->, thick, dashed, black!40] (user-left) |- node[near end, above, font=\scriptsize, black!50] {Skips} (ca-left.west);

\draw[decorate, decoration={brace, amplitude=5pt}, thick, red!60!black]
    (-1.1, 1.7) -- (-1.1, -4.6) node[midway, right=8pt, font=\scriptsize, text width=2.2cm, align=left, red!60!black] {Substantive content\\not reached};

\node[label] at (4.2, 5.2) {Proposed Standard};
\node[font=\scriptsize\itshape] at (4.2, 4.8) {Universal Substantive Disclosure};

\node[section, minimum height=3.4cm] (gen-right) at (4.2, 2.6) {};
\node[font=\footnotesize\bfseries] at (4.2, 4.0) {General Sections};
\node[subsection, substantive] at (4.2, 3.2) {Data sales disclosure};
\node[subsection, substantive] at (4.2, 2.5) {Automated profiling};
\node[subsection, substantive] at (4.2, 1.8) {Biometric collection};

\node[section, minimum height=1.3cm] (ca-right) at (4.2, -0.4) {};
\node[font=\footnotesize\bfseries] at (4.2, 0.0) {California Privacy Rights};
\node[subsection, procedural] at (4.2, -0.65) {Opt-out mechanism};

\node[section, minimum height=1.3cm] (eu-right) at (4.2, -2.0) {};
\node[font=\footnotesize\bfseries] at (4.2, -1.6) {EU/UK Users};
\node[subsection, procedural] at (4.2, -2.25) {Data subject rights};

\node[section, minimum height=1.3cm] (il-right) at (4.2, -3.5) {};
\node[font=\footnotesize\bfseries] at (4.2, -3.1) {Illinois Residents};
\node[subsection, procedural] at (4.2, -3.75) {Written consent};

\node[user] (user-right) at (8.2, 4.8) {User};
\node[note, text width=2cm] at (8.2, 4.1) {(Texas resident)};

\draw[->, thick, black] (user-right.south west) -- node[above left=-1pt, font=\scriptsize] {Reads} (gen-right.north east);

\draw[decorate, decoration={brace, amplitude=5pt}, thick, black!60!green]
    (7.3, 3.5) -- (7.3, 1.5) node[midway, right=8pt, font=\scriptsize, text width=2.5cm, align=left, black!60!green] {Substantive content\\accessible};

\node[subsection, substantive, minimum width=1.5cm] (leg-s) at (-1.5, -5.3) {};
\node[font=\scriptsize, right=3pt of leg-s] {Substantive disclosure};
\node[subsection, procedural, minimum width=1.5cm] (leg-p) at (3.0, -5.3) {};
\node[font=\scriptsize, right=3pt of leg-p] {Procedural content};
\node[subsection, vague, minimum width=1.5cm] (leg-v) at (-5.5, -5.3) {};
\node[font=\scriptsize, right=3pt of leg-v] {Vague/hedged language};

\end{tikzpicture}%
}
\caption{Jurisdiction-siloed disclosure (left) versus universal substantive disclosure (right): illustrative comparison. In the current pattern, substantive disclosures about data practices appear only within jurisdiction-specific sections, which users outside those jurisdictions are predicted to skip based on information foraging theory. Under the proposed standard, substantive content appears in the main policy body accessible to all users, while jurisdiction-specific sections contain only procedural information about exercising rights. Abstract labels represent categories of disclosure, not prescribed policy text.}
\label{fig:structural_comparison}
\end{figure*}
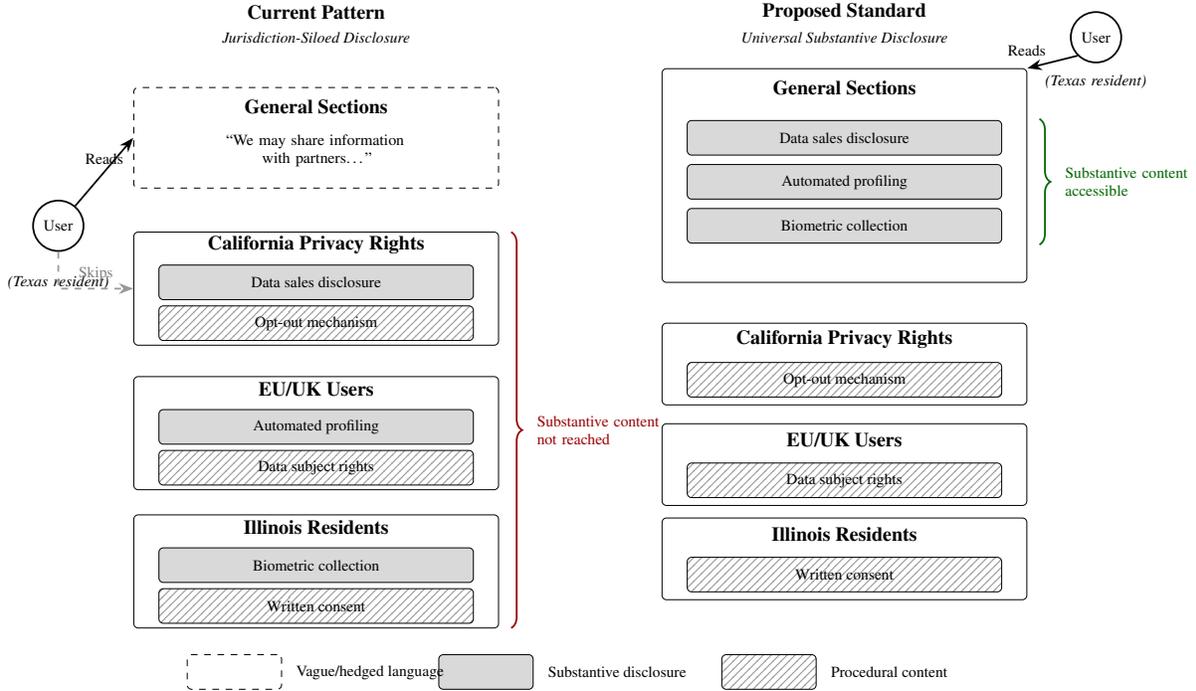

\section{Methodology}
\label{sec:methodology}

\subsection{Company Selection}

We audited privacy policies from 123 companies across 14 industries, selected to represent major platforms whose policies affect large user populations:

\begin{itemize}
    \item Major technology platforms (Google, Meta, Amazon, Microsoft, Apple)
    \item AI/ML companies (OpenAI, Anthropic, Midjourney, Stability AI)
    \item Financial services (Chase, PayPal, Coinbase, Robinhood)
    \item Healthcare and wellness (23andMe, BetterHelp, Cerebral, CVS)
    \item Surveillance and defense (Palantir, Clearview AI, Anduril, Cellebrite)
    \item Data brokers (Gravy Analytics, LexisNexis, Experian)
    \item Social and dating platforms (Tinder, Bumble, Grindr, LinkedIn)
    \item Travel, gaming, and e-commerce
\end{itemize}

This is a purposive sample of major companies, not a random sample of the web. Findings about prevalence rates apply to significant platforms, not websites generally. Our 123-company sample represents what Srinath et al.~\cite{srinath2021privaseer} identify as more popular domains by PageRank value: large, legally sophisticated, multinational organizations subject to regulatory scrutiny across multiple jurisdictions. The PrivaSeer corpus of over one million privacy policies demonstrates that such organizations systematically differ from typical websites: their policies are longer, cover more topics, and address more regulatory frameworks than the web-scale population. Our sample's mean policy length of approximately 4,200 words substantially exceeds the PrivaSeer corpus mean of 1,871 words, consistent with policies from high-traffic, multinational organizations.

This sampling strategy is intentional. Jurisdiction-siloed disclosure is most likely to emerge at organizations with (a) legal resources to create jurisdiction-specific sections, (b) multinational operations triggering multiple regulatory regimes, and (c) user bases large enough to warrant compliance investment. Smaller organizations, comprising the majority of the PrivaSeer corpus, may lack jurisdiction-specific sections entirely, not because they practice universal disclosure, but because they lack resources for jurisdiction-specific compliance. Our sample targets the stratum where the phenomenon can meaningfully be documented and where the impact is greatest: the platforms that collectively reach billions of users.

\subsection{Policy Collection}

We collected policies in January 2026 using native HTTP requests, browser automation for JavaScript-rendered sites, and the Internet Archive for sites blocking automated access. We collected a single privacy policy per company, the primary English-language version accessible from each company's main domain (e.g., company.com). We did not systematically collect regional subsidiary policies that may exist on country-specific domains (e.g., company.eu, company.co.uk, company.de), though some companies in our sample explicitly reference such policies (see Limitations). This methodological choice means our findings are most directly applicable to users encountering the primary policy version.

Our analysis accessed policies via US-based IP addresses and examined primary domain policies. Some multinational companies maintain regional subsidiary policies (e.g., company.eu, company.co.uk) that may present substantive disclosures more prominently to users in those jurisdictions. Our findings should be interpreted as reflecting the experience of users accessing global policies rather than region-specific variants. This limitation is common in privacy policy research; most existing corpora, including OPP-115 and PrivaSeer, similarly focus on primary English-language policies~\cite{wilson2016creation,srinath2021privaseer}.

\subsection{Segmentation and Classification}
\label{sec:classification}

Policies were segmented using HTML heading structure, preserving author-intended organization. Each segment was classified using a 14-category taxonomy derived from OPP-115~\cite{wilson2016creation} and extended to address post-2018 regulatory developments (see Appendix B for complete category definitions). The six categories relevant for identifying jurisdiction-siloed disclosure are:

\begin{itemize}
    \item REGIONAL: Jurisdiction-specific rights procedures (California, EU/UK, Illinois sections)
    \item SALE\_SHARING: Disclosures about data sales or sharing for advertising
    \item SENSITIVE\_DATA: Biometric, health, genetic, or other sensitive data practices
    \item AUTOMATED\_DECISIONS: Profiling, algorithmic decisions, AI-driven processing
    \item FIRST\_PARTY: Direct data collection and use
    \item THIRD\_PARTY: Data sharing with external parties
\end{itemize}

Classification was performed using three frontier LLMs from different providers---Claude Haiku 4.5 (Anthropic), GPT-5.2 (OpenAI), and Gemini-3-flash (Google)---with a shared classification prompt developed by the author, a certified privacy professional (CIPP/US, AIGP) with five years of experience in enterprise data governance and compliance architecture at Microsoft, where he served as AI/ML Compliance SME and led the M365 Trusted Platform Compliance Assurance Program. The classification prompt was iteratively refined during annotation of the full 123-policy corpus: initial prompt versions achieved approximately 78\% inter-model agreement on complex policies, while the final prompt achieved 93\%+ agreement. This refinement process resolved approximately 280 classification disputes through human adjudication during iterative prompt development and produced eight documented boundary distinctions (e.g., distinguishing AI-related platitudes from substantive automated decision disclosures; separating jurisdiction-specific procedures from substantive rights). These boundary rules are documented in Appendix~B.

This three-model methodology provides some protection against idiosyncratic model behavior, though models from different providers likely share training data and may exhibit similar limitations with legal text. Each segment received independent classifications from all three models, with final labels determined by majority vote. This achieved 78.3\% unanimous agreement (all three models agree) and 98.7\% usable consensus (at least two models agree). Pairwise agreement rates were: Claude$\leftrightarrow$Gemini 85.5\%, GPT$\leftrightarrow$Gemini 86.1\%, Claude$\leftrightarrow$GPT 83.8\%.

We report three distinct agreement metrics that serve different purposes and should not be conflated: (1) \emph{raw agreement percentages} (78.3\% unanimous, 98.7\% majority) measure classification consistency (whether models produce the same output); (2) \emph{Fleiss' Kappa} (0.858, reported below) is a chance-corrected statistic that measures whether agreement exceeds what random classification would produce; and (3) \emph{Cohen's Kappa against OPP-115} (0.629, also below) measures validity (whether our classifications match an independent human standard). High percentage agreement demonstrates reproducibility; Kappa demonstrates this agreement is meaningful; OPP-115 validation demonstrates alignment with human judgment.

We report these as \emph{consistency} metrics reflecting classification reproducibility across different model architectures, not as ground-truth accuracy validated against independent human annotators. Importantly, prompt refinement occurred during annotation of the full corpus rather than on a held-out validation sample; the final prompt and boundary rules were developed through continuous iteration on the same policies used for analysis. This approach prioritized consistent application of expert judgment across the entire dataset but means the methodology cannot claim validation against independent data. The classification serves as a tool for identifying disclosure patterns rather than as a benchmark comparable to multi-annotator corpora like OPP-115~\cite{wilson2016creation}, which used three law students annotating independently with formal inter-annotator reliability measurement.

\subsubsection{Methodological Comparison with OPP-115}
\label{sec:opp115_comparison}

Our annotation methodology differs from OPP-115 in ways that warrant explicit discussion. OPP-115 employed three law students who annotated independently using a shared codebook, with inter-annotator reliability measured via Fleiss' Kappa~\cite{wilson2016creation}. Our approach uses three LLMs from different providers operating on a shared classification prompt developed by a domain expert.

This approach builds upon emerging practices in computational text analysis. Gilardi et al.~\cite{gilardi2023chatgpt} demonstrated that LLMs can match or exceed crowd-worker accuracy for annotation tasks using simple codebook-derived prompts. We extend this foundation with (1) a highly engineered classification prompt iteratively refined by a domain expert to align with privacy law expertise, (2) three models from different providers to reduce single-model artifacts, and (3) validation against OPP-115's human expert annotations. While frontier LLMs may align with human consensus on well-specified tasks, this likely reflects prompt clarity and task tractability rather than independent legal judgment, a limitation we address through explicit validation against human-annotated ground truth.

These approaches represent different tradeoffs. Human annotators bring legal training and contextual judgment that may recognize subtle distinctions, but also exhibit substantial disagreement: OPP-115 reports Fleiss' Kappa values as low as 0.49 for ambiguous categories~\cite{wilson2016creation}. LLM annotators execute instructions more consistently but cannot exercise independent legal judgment; their agreement reflects prompt clarity and task tractability rather than convergent expert opinion.

Critically, three-model agreement should not be interpreted as equivalent to human inter-annotator agreement. Models from different providers likely share substantial training data and may exhibit correlated limitations with legal text. The 78.3\% unanimous agreement demonstrates that the classification task is \emph{reproducible} across different architectures, an important property for scientific replication, but does not establish ground-truth validity in the way that convergent expert judgment would. Fleiss' Kappa among the three models was 0.858, indicating ``almost perfect agreement''~\cite{landis1977measurement}, but this reflects prompt clarity and task tractability rather than convergent expert judgment. Notably, among the 38 three-way disagreements, REGIONAL appeared as a disputed category in 47.4\% of cases, making the substantive/procedural boundary the most contested classification decision.

To assess validity, we validated our methodology against OPP-115's human annotations. Classifying all 3,726 OPP-115 segments with our prompt yields Cohen's Kappa of 0.629 against the human majority-vote labels, falling within the range of per-category Fleiss' Kappa values (0.49--0.91, mean 0.71) reported among the original human annotators~\cite{wilson2016creation}. On segments where all three OPP-115 annotators agreed unanimously, our accuracy reaches 81.3\%; on disputed segments, accuracy drops to 57.7\%, reflecting that our model struggles precisely where human experts struggled. This pattern suggests our methodology captures the difficulty structure of the classification task rather than introducing systematic bias.

\subsubsection{The Circularity Problem}
\label{sec:circularity}

A critical methodological limitation must be stated explicitly: our classification rules were developed on the same corpus used for analysis. This creates potential circularity: rules may conform to the data rather than the data confirming pre-existing rules.

Specifically:
\begin{itemize}
    \item The classification prompt underwent iterative refinement during annotation of all 123 policies
    \item The eight boundary distinctions in Appendix~B emerged from resolving disputes in this corpus
    \item No held-out validation sample was used to test classification rules before applying them to the analysis corpus
\end{itemize}

This approach prioritized consistent application of expert judgment across the dataset but means the prevalence estimate (62.6\%) and instance count (282) cannot be treated as validated measurements. They represent what our methodology detects when applied to this corpus, not necessarily what independent methods would detect or what exists in the broader population.

To address potential confirmation bias, we examined whether dispute resolutions systematically favored findings of jurisdiction-siloed disclosure. Of the approximately 280 classification disputes resolved during iterative prompt refinement, we tracked resolution direction for a subset: 35 disputes were resolved to substantive categories that could support siloed findings (FIRST\_PARTY, THIRD\_PARTY, SALE\_SHARING, SENSITIVE\_DATA, AUTOMATED\_DECISIONS), while 56 were resolved to procedural/structural categories that would not support siloed findings (REGIONAL, OTHER, USER\_CHOICE, USER\_ACCESS, INTL\_SPECIFIC). This distribution does not reveal systematic bias toward detecting siloed disclosure. Additionally, 93.6\% (264 of 282) of our siloed disclosures contain explicit substantive language (e.g., ``We sell,'' ``We collect'') rather than implied disclosures inferred from procedural rights language, suggesting our classifications capture genuine substantive information rather than over-inferences from ambiguous text.

We report three-model agreement metrics (78.3\% unanimous) as measures of \emph{reproducibility} (can different models produce the same classifications?) rather than \emph{validity} (are these classifications correct?). This distinction is critical and should not be conflated. The OPP-115 validation described above provides direct external validation for the six pre-2016 categories in our taxonomy (FIRST\_PARTY, THIRD\_PARTY, USER\_CHOICE, etc.). However, it cannot validate the six post-2018 categories (REGIONAL, SALE\_SHARING, AUTOMATED\_DECISIONS, SENSITIVE\_DATA, INTL\_SPECIFIC) that did not exist in the 2016 taxonomy, precisely the categories most central to our claims about jurisdiction-siloed disclosure.

Consequently, we present findings about post-2018 categories as preliminary observations requiring independent validation. The 69 SALE\_SHARING instances, 41 SENSITIVE\_DATA instances, and 34 AUTOMATED\_DECISIONS instances classified as jurisdiction-siloed rest on internally-consistent classification but lack independent human validation. Our overall prevalence estimate combines both validated and unvalidated categories; a conservative estimate accounting only for validated pre-2016 categories would encompass approximately 138 instances (49\% of identified cases), still supporting the hypothesis that jurisdiction-siloed disclosure is prevalent but providing more modest magnitude. The pattern-level claim, that companies disclose practices exclusively in jurisdiction-specific sections, remains robust because it rests on the formal logic of the definition rather than category-level correctness.

This limitation is why we frame this work as hypothesis-generating and release our methodology for independent validation.

\subsubsection{External Validation Efforts}

To assess validity against independent standards, we pursued multiple external validations. First, the OPP-115 validation described above provides external validation against an independently annotated corpus, with accuracy reaching 81.3\% on segments where all three human annotators agreed unanimously.

Second, we examined whether domain-specific models trained on privacy policy corpora could serve as validation baselines. PrivBERT~\cite{srinath2021privaseer}, pretrained on over one million privacy policies, achieves only 5.1\% agreement with our classifications ($\kappa = -0.038$, worse than chance), despite achieving 74\% accuracy on the OPP-115 test set. Critically, this failure does not validate our methodology---an obsolete tool's inability to classify contemporary policies does not establish that our approach is correct. Rather, it reveals that temporal domain shift has rendered pre-2018 models obsolete for contemporary analysis: PrivBERT lacks training data for post-2018 categories (REGIONAL, SALE\_SHARING, AUTOMATED\_DECISIONS, SENSITIVE\_DATA) central to modern policies. This field-wide validation gap---the most rigorous human-annotated corpus predates the current regulatory landscape---is why we distinguish between validated pre-2016 categories (49\% of instances) and preliminary post-2018 categories (51\% of instances).

Third, comparative fine-tuning experiments provide indirect validation that our taxonomy refinements improved classification coherence. Training PrivBERT on our OPPT corpus achieves 82.1\% accuracy ($\pm$1.6\%) versus 74.4\% ($\pm$0.6\%) when trained on OPP-115, a statistically significant improvement ($t=9.927$, $p<0.00001$) across five random seeds. This does not establish that our classifications are correct, but it demonstrates that the boundary distinctions developed through iterative refinement (e.g., distinguishing REGIONAL from INTL\_SPECIFIC, separating substantive from procedural content) produce more learnable categories than OPP-115's original taxonomy.

Our contribution is not a benchmark classification methodology but rather a tool for identifying disclosure patterns. The pattern-level finding, that jurisdiction-siloed disclosure is prevalent, is robust to minor classification variations, and the validation portfolio (OPP-115 human agreement, PrivBERT temporal analysis, fine-tuning learnability) collectively supports the methodology's fitness for this purpose. This multi-method validation approach reflects emerging best practices in NLP research when facing temporal domain shift: rather than relying on a single potentially obsolete benchmark, we combine complementary evidence streams that collectively strengthen confidence in the methodology's fitness for purpose.

\subsection{Identifying Jurisdiction-Siloed Disclosure}

We operationalize jurisdiction-siloed disclosure as follows: a disclosure is siloed if substantive content (SALE\_SHARING, SENSITIVE\_DATA, AUTOMATED\_DECISIONS, FIRST\_PARTY, or THIRD\_PARTY) appears within a REGIONAL section \emph{without equivalent disclosure in a universal section}.

This is a conservative definition. Companies that disclose practices both universally and in regional sections are not flagged. We identify only cases where the \emph{only} disclosure of a substantive practice appears in a jurisdiction-specific section.

\subsection{Operationalizing Equivalent Disclosure}
\label{sec:equivalence}

A critical methodological question is what constitutes ``equivalent disclosure'' in the universal sections. We adopt a semantic equivalence standard: universal disclosure must communicate the same practice, not merely related or adjacent information.

\subsubsection{Coding Rules for Equivalence}

We apply three criteria to determine whether universal disclosure is equivalent to jurisdiction-specific disclosure:

\begin{enumerate}
    \item Practice identity: The universal section must disclose the \emph{same} practice, not a broader category that might or might not include it. ``We share data with partners'' is not equivalent to ``We sell your personal information'' because sharing encompasses practices (joint ventures, service providers) that are not sales.

    \item Specificity preservation: If the regional disclosure is specific (``We collect facial geometry''), a vague universal disclosure (``We may collect information about you'') is not equivalent. The universal disclosure must match or exceed the specificity of the regional disclosure.

    \item Semantic clarity: The universal disclosure must use language that would communicate the practice to a reasonable reader. Euphemisms or obfuscating language that technically encompasses a practice but would not alert readers to it do not constitute equivalence.
\end{enumerate}

\subsubsection{Worked Examples}

We illustrate the application of these criteria with three borderline cases encountered during coding.

\emph{Example 1: Sale vs.\ Sharing (Not Equivalent).}

\textit{California section}: ``We sell your personal information to third parties for monetary consideration.''

\textit{Main body}: ``We may share your information with our advertising partners to deliver relevant ads.''

\textit{Coding decision}: NOT equivalent. Under CCPA, ``sale'' has a specific legal meaning requiring exchange for valuable consideration. ``Sharing with advertising partners'' could describe contextual advertising, first-party ad targeting, or practices that do not constitute sales. A reader of the main body would not learn that monetary transactions for their data occur. The California disclosure is coded as siloed.

\textit{Corrected universal disclosure}: A policy satisfying universal substantive disclosure would state in the main body: ``We sell your personal information to third parties for monetary consideration. We also share information with advertising partners to deliver relevant ads.'' The California section would then contain only procedural content: ``California residents have the right to opt out of the sale of their personal information by [mechanism].'' This restructuring preserves compliance while ensuring all users learn about data sales.

\emph{Example 2: Biometric Collection (Not Equivalent).}

\textit{Illinois section}: ``We collect biometric identifiers, including facial geometry extracted from photographs you upload.''

\textit{Main body}: ``We use facial recognition technology to help you tag friends in photos.''

\textit{Coding decision}: NOT equivalent. While both reference facial recognition, the main body frames the practice as a user-facing feature rather than disclosing biometric data collection. A reader might understand that facial recognition occurs without understanding that biometric identifiers are collected, stored, and potentially shared. The substantive disclosure---that biometric identifiers are collected---appears only in the Illinois section and is coded as siloed.

\textit{Corrected universal disclosure}: A compliant policy would state in the main body: ``We collect biometric identifiers, including facial geometry extracted from photographs you upload. We use this information to power facial recognition features that help you tag friends in photos.'' The Illinois section would contain only: ``Illinois residents may request deletion of biometric identifiers under BIPA by [mechanism].'' The substantive practice---biometric collection---becomes universally visible; only the exercise mechanism remains jurisdiction-specific.

\emph{Example 3: Automated Decision-Making (Equivalent).}

\textit{EU section}: ``You have the right not to be subject to decisions based solely on automated processing, including profiling, which produces legal effects concerning you.''

\textit{Main body}: ``We use automated systems to make decisions about your eligibility for certain products and services. These automated decisions may affect your access to credit, insurance, or employment opportunities.''

\textit{Coding decision}: Equivalent. The main body discloses that automated decision-making occurs and that it affects consequential outcomes. The EU section adds procedural information (the right to contest) but does not disclose a practice absent from the main body. This is NOT coded as siloed because the substantive practice---automated decision-making with material effects---is universally disclosed.

\subsubsection{Special Coding Rules}

Several edge cases required explicit coding rules.

When a California section states ``Right to opt out of sale,'' this implies sales occur. We code such implied practices as substantive disclosures if no explicit universal disclosure exists, because a user encountering this statement learns about the practice, demonstrating that the knowledge is siloed even when expressed procedurally.

Universal disclosure of a category (``We collect sensitive personal information'') may or may not be equivalent to specific regional disclosures (``We collect precise geolocation data''). We apply a specificity criterion: if the universal category is sufficiently detailed that a reasonable reader would understand the specific practice is included, we treat it as equivalent. Bare categorical disclosure without enumeration is not equivalent to specific regional disclosures.

Universal disclosures frequently use hedged language (``We may sell,'' ``We might share''). We treat hedged universal disclosures as equivalent to categorical regional disclosures (``We sell'') because both communicate that the practice can occur. However, we note that hedged language may itself constitute a transparency problem beyond the scope of this study.

The 47 cases (1.3\% of segments) where the three-model consensus did not produce majority agreement on equivalence were resolved by the author applying these criteria. In 31 cases, the author determined equivalence existed; in 16 cases, the disclosure was coded as siloed.

\subsection{Limitations}

Several limitations warrant acknowledgment regarding sample generalizability, classification methodology, and our universality assumption.

Our prevalence finding applies specifically to major platforms and should not be extrapolated to websites generally. The PrivaSeer corpus provides useful context for understanding what our sample represents within the broader population. That corpus contains over one million English-language privacy policies with substantial diversity: 63\% from .com domains, policies ranging from 143 to nearly 17,000 words, and Flesch-Kincaid readability scores requiring 14.87 years of education on average. Critically, PrivaSeer's analysis found that higher-PageRank domains, organizations comparable to those in our sample, systematically differ from the corpus mean: they maintain longer policies, address more topics, and face greater regulatory complexity~\cite{srinath2021privaseer}.

This suggests our sample represents a specific stratum: the upper tail of organizational sophistication and regulatory exposure. Whether jurisdiction-siloed disclosure occurs at similar rates among smaller organizations remains an open empirical question. Two competing hypotheses are plausible: (1) smaller organizations may exhibit lower rates because they lack the legal resources to create jurisdiction-specific sections, resulting in shorter, undifferentiated policies; or (2) smaller organizations may exhibit similar or higher rates if they adopt templates from larger competitors or compliance vendors that embed jurisdiction-specific structures. Future research using web-scale corpora could test these hypotheses.

These findings are transferable to organizations of similar scale, legal sophistication, and multinational regulatory exposure. The extent to which smaller companies, international firms outside Western regulatory frameworks, or specific industries exhibit similar patterns remains an open empirical question requiring dedicated sampling strategies. Future research should examine whether jurisdiction-siloed disclosure appears at comparable rates in these populations.

Our sample's value lies not in population representativeness but in documenting the phenomenon where it matters most: the platforms with the largest user bases, the most sophisticated legal operations, and the greatest regulatory scrutiny. If jurisdiction-siloed disclosure undermines transparency at Google, Meta, and Microsoft, the impact on informed consent is substantial regardless of prevalence rates among smaller websites.

The 18 companies in our sample (14.6\%) that maintain no jurisdiction-specific sections warrant consideration for what they reveal about alternative disclosure strategies, and whether avoiding regional sections correlates with better transparency.

We compared disclosure comprehensiveness across three groups: companies without regional sections (18), companies with appropriately-used regional sections containing only procedural content (28), and companies with jurisdiction-siloed substantive disclosures (77). Measuring coverage across five major practice categories (first-party collection, third-party sharing, sale/sharing, sensitive data, automated decisions), we find that companies \emph{without} regional sections disclose fewer categories on average (3.94 of 5) than companies with appropriately-used regional sections (4.04 of 5). Only 33.3\% of no-regional-section companies disclose all five major categories, compared to 42.9\% of companies using regional sections appropriately.

This finding directly addresses a potential alternative interpretation: that avoiding regional sections signals a commitment to universal disclosure. The data suggest otherwise. Companies without regional sections include (1) smaller or regionally-focused organizations that may lack the legal resources or multinational exposure to require jurisdiction-specific compliance infrastructure; (2) companies in regulated industries where sector-specific regulations already mandate uniform disclosure; and (3) a genuine subset practicing comprehensive universal disclosure (e.g., Adobe, TikTok, and Discord, all disclosing 5/5 major categories). However, this third pattern accounts for only one-third of no-regional-section companies. The remainder practice what we term ``regulatory minimalism'': disclosing little to anyone rather than disclosing comprehensively to all. The absence of regional sections does not automatically indicate transparency virtue; some companies simply have less to disclose or disclose less regardless of organization.

Notably, companies \emph{with} jurisdiction-siloed disclosures paradoxically show the highest category coverage (4.65 of 5, with 71.4\% disclosing all five categories). This is not a transparency success; these companies achieve comprehensiveness through organizational siloing, with substantive practices placed in jurisdiction-specific sections where, based on information foraging research, users outside those jurisdictions may be unlikely to encounter them. The comprehensiveness is measured at the policy level, not the user experience level.

The classification methodology reflects alignment with a single expert's professional judgment encoded in a shared prompt, not consensus among multiple independent human annotators. A different expert, particularly one with different professional experience (e.g., academic rather than industry background), might produce different classification guidelines. We report three-model agreement (78.3\% unanimous, 98.7\% majority) as a consistency metric demonstrating that the classification task produces reproducible results across different LLM architectures, but this measures prompt clarity and task tractability, not ground-truth accuracy against an external standard. The pattern-level findings (jurisdiction-siloed disclosure) are robust to minor classification differences; the specific counts should be interpreted as estimates rather than precise measurements.

Policies were collected in January 2026 and may have since changed. Privacy policies are living documents that companies update in response to regulatory developments and business changes.

Finally, our analysis collected only a single privacy policy per company, typically the primary English-language version accessible from the company's main domain (e.g., company.com). We did not systematically collect or analyze regional subsidiary policies that may exist on country-specific domains (e.g., company.eu, company.co.uk, company.de). Several companies in our sample explicitly reference such regional policies: Wise notes that ``Wise Europe SA'' and ``Wise Payments Limited'' maintain separate privacy documentation; Zoom references jurisdiction-specific privacy statements; Microsoft describes regional data handling in supplementary documents. This methodological choice has important implications for interpreting our findings. Some disclosures we classify as ``jurisdiction-siloed'' within a single global policy might correspond to practices that are disclosed more prominently in a separate regional subsidiary policy; a user accessing Microsoft's EU subsidiary site, for instance, might encounter different or more detailed disclosures than those in the global policy we analyzed. Our study cannot distinguish between practices that are (a) disclosed only in a jurisdiction-specific section of a single global policy, (b) disclosed in a separate regional subsidiary policy we did not collect, or (c) not disclosed at all outside the jurisdiction-specific section. This limitation means our findings are most directly applicable to users who encounter the primary English-language policy, likely the majority of users, but not all. Future work should systematically compare global and regional subsidiary policies to determine whether jurisdiction-siloed disclosures in global policies are remedied by more prominent disclosure in regional versions, or whether the concealment effect persists across a company's entire policy ecosystem.

Our analysis rests on a universality assumption: that practices disclosed in jurisdiction-specific sections affect all users---that when Clearview AI discloses biometric collection in an Illinois section, they collect biometrics from users worldwide; that when Thomson Reuters discloses data sales to Californians, they sell data from users in all states. This assumption, while reasonable for many practices, cannot be verified from policy text alone and represents a limitation of our methodology.

Some practices genuinely vary by jurisdiction for legitimate reasons. Data transfers may differ based on GDPR adequacy decisions; a company might transfer EU user data only to countries with adequate data protection, while transferring US user data more broadly. Automated decision-making might be used only for EU applicants to comply with Article~22 transparency requirements, while US applicants are processed differently. Data retention timelines could legitimately differ across jurisdictions to comply with varying legal requirements. Regional infrastructure differences (EU data centers, US data centers) may result in genuinely different processing activities. Legal bases for processing (consent in the EU, legitimate interests elsewhere) may correspond to substantively different data practices, not merely different legal justifications for identical practices.

Our methodology cannot distinguish between two scenarios: (1) a company engages in a universal practice but discloses it only where legally required (the transparency limitation we identify), and (2) a company engages in a practice only in certain jurisdictions and appropriately discloses it only there (legitimate regional variation). The former represents a transparency failure; the latter represents appropriate disclosure.

Several factors support the concealment interpretation, though both interpretations remain plausible without technical verification. Biometric databases like Clearview's contain faces regardless of the subject's residence; data broker sales do not typically exclude non-California data; ad personalization operates on users regardless of location. Technical infrastructure makes jurisdiction-specific processing costly; it is simpler to process all data identically and vary only which users receive disclosure. Moreover, if practices genuinely varied by jurisdiction, we would expect some indication in the policy text (``We do not sell data from non-California residents''), yet such clarifications are absent. These observations are consistent with concealment, though regulatory inspection of specific companies would be needed to confirm.

Future work could empirically test the universality assumption through technical analysis (examining whether identical tracking mechanisms operate across jurisdictions), user surveys (asking whether users in different jurisdictions experience different data practices), or regulatory enforcement records (identifying cases where companies claimed jurisdiction-specific practices but were found to operate universally). A framework for distinguishing legitimate regional variation from disclosure concealment might consider: (1) whether technical infrastructure plausibly supports jurisdiction-specific processing, (2) whether the practice type inherently varies by jurisdiction (e.g., legal basis) or is jurisdiction-independent (e.g., biometric collection), and (3) whether the policy explicitly states that non-disclosed jurisdictions are treated differently. Several features of our findings support the concealment interpretation over legitimate regional variation. First, 264 of 282 siloed disclosures (93.6\%) contain explicit substantive statements, direct assertions like ``We sell your personal information'' or ``We collect biometric identifiers,'' rather than mere procedural language from which practices might be inferred. This explicit/implied ratio suggests companies are withholding material information, not merely organizing procedural rights by jurisdiction.

Second, certain practice categories show patterns inconsistent with legitimate regional variation. International data transfer disclosures (FIRST\_PARTY and THIRD\_PARTY appearing in international-specific sections) account for 49 instances, all explicit. Data collection and cross-border transfer mechanisms typically operate uniformly across a company's infrastructure; a company that transfers EU user data to third countries almost certainly transfers US user data as well, yet discloses the practice only where GDPR requires it. Similarly, the 34 automated decision-making disclosures appear exclusively in GDPR sections, despite the underlying algorithmic systems operating identically for users worldwide.

Third, platform companies with globally unified infrastructure concentrate siloed disclosures: Roblox (41 instances), PayPal (11), Verizon (10), and LinkedIn (9). These companies operate identical technical systems across jurisdictions, making jurisdiction-specific processing implausible. Roblox's investor disclosures confirm over 70 million daily active users globally on shared infrastructure, yet its EU sections contain 41 substantive disclosures absent from universal sections. PayPal's SEC filings document a unified global payments platform processing transactions identically regardless of user location; Verizon's annual reports describe integrated network infrastructure serving customers across all operating regions; LinkedIn's engineering documentation confirms a single global platform architecture. For these companies, the technical infrastructure required to process users differently by jurisdiction would represent a substantial engineering investment with no documented business justification.

Fourth, explicit geographic limitation language is notably absent. If practices genuinely varied by jurisdiction, we would expect clarifying statements (``This practice applies only to California residents'' or ``We do not sell data from users outside California''). We found no such limitations in our corpus. Notably, when companies \emph{do} wish to communicate geographic restrictions, the language exists and is used: we observed statements like ``This service is available only in the United States'' for product availability and ``EU user data is processed exclusively in our European data centers'' for infrastructure limitations. The contrast is instructive: companies employ explicit geographic limitation language when they intend it, suggesting that its absence from substantive disclosures reflects universal practices rather than unstated limitations. Companies that genuinely restrict practices geographically could overcome the concealment inference by stating so explicitly; their silence suggests the practices are universal.

These patterns do not definitively establish universality for all 282 instances. Some disclosures, particularly those involving data retention timelines or regional infrastructure, may reflect legitimate variation. We therefore propose a tiered inference framework, summarized in Table~\ref{tab:confidence_tiers}. This framework provides transparency about which findings rest on strong evidence versus reasonable inference.

The majority of instances (62.1\%) fall into Strongly or Moderately Inferred tiers, reflecting practices (international data transfers, automated decision-making, biometric collection) that rarely vary by jurisdiction due to technical infrastructure constraints. The 107 Weakly Inferred instances (37.9\%) involve practice types where legitimate regional variation is more plausible: data retention disclosures (31 instances, 29\% of Weakly Inferred), where jurisdictions impose different minimum retention periods; regional processing arrangements (24 instances, 22\%), where companies may maintain separate infrastructure; and sale/sharing disclosures (52 instances, 49\%), where geographic limitation of data broker relationships is technically feasible though uncommon. Even within this tier, the absence of explicit geographic limitation language in our corpus suggests concealment remains the more likely interpretation. Future work should empirically validate each tier. Until such validation exists, our findings should be interpreted as identifying \emph{potential} concealment patterns that warrant scrutiny, with confidence levels indicating the strength of the universality inference.

\begin{table}[t]
\caption{Universality inference tiers for jurisdiction-siloed disclosure instances}
\label{tab:confidence_tiers}
\small
\begin{tabular}{@{}lrr>{\raggedright\arraybackslash}p{4.8cm}@{}}
\toprule
\textbf{Inference Tier} & \textbf{Count} & \textbf{\%} & \textbf{Justification} \\
\midrule
Verified & 1 & 0.4\% & External documentation directly confirming disclosed practice applies globally (Clearview AI) \\
\midrule
Strongly Inferred & 77 & 27.3\% & International transfers (49) + platform companies with documented global infrastructure where regional variation is implausible (28) \\
\midrule
Moderately Inferred & 97 & 34.4\% & Automated decisions (34), biometric/sensitive data (38), foundational collection practices (25) \\
\midrule
Weakly Inferred & 107 & 37.9\% & Data retention, regional processing, sale/sharing where geographic limitation is plausible \\
\midrule
\textbf{Total} & \textbf{282} & \textbf{100\%} & \\
\bottomrule
\end{tabular}
\end{table}

\FloatBarrier
\section{Findings}

The findings below document \emph{organizational patterns}---where disclosures appear within policy structure---rather than user behavior. The transparency implications rest on the behavioral assumption discussed in Section~\ref{sec:methodology} and requiring the validation described in Section~\ref{sec:behavioral_validation}.

\subsection{Prevalence of Jurisdiction-Siloed Disclosure}
\label{sec:prevalence}

A conservative estimate of jurisdiction-siloed disclosure, restricted to the two pre-2016 categories validated against OPP-115 human annotations (Cohen's $\kappa = 0.629$)---FIRST\_PARTY\_COLLECTION and THIRD\_PARTY\_SHARING---identifies 138 instances across 54 companies (44\% of sample) where substantive disclosures appear only within jurisdiction-specific sections (see Section~\ref{sec:equivalence} for equivalence criteria). Our full methodology, extending to post-2018 categories (SALE\_SHARING, SENSITIVE\_DATA, AUTOMATED\_DECISIONS) that await independent human validation (see Section~\ref{sec:opp115_comparison}), identifies 282 potential instances across 77 companies (62.6\% of sample). The pattern-level finding that jurisdiction-siloed disclosure is prevalent remains robust under both estimates; only the magnitude varies.

Of these 282 instances, 264 (93.6\%) contain explicit substantive statements (``We sell your personal information,'' ``We collect biometric identifiers''); only 18 instances (6.4\%) rely on inference from procedural language (e.g., an opt-out right implying the underlying practice). These counts are preliminary and require independent validation as discussed in Section~\ref{sec:opp115_comparison}.

Reflecting our inference framework (Table~\ref{tab:confidence_tiers}), 175 instances (62.1\%) fall into Strongly or Moderately Inferred tiers, practices where legitimate regional variation is implausible, including international data transfers, automated decision-making, and biometric collection. The remaining 107 instances (37.9\%) involve data retention, regional processing, or sale/sharing where geographic limitation is technically feasible. Because individual companies can have instances across multiple inference tiers, company-level statistics require careful interpretation. Only 1 company (Clearview AI, 0.8\% of sample) has externally verified disclosure practices. Of the 77 companies exhibiting jurisdiction-siloed disclosure, 45 (58.4\% of affected companies; 36.6\% of full sample) have at least one instance in a Strongly or Moderately Inferred tier, meaning more than half of affected companies exhibit siloing for practices where legitimate regional variation is implausible. The remaining 31 affected companies (40.3\% of affected; 25.2\% of sample) have instances only in the Weakly Inferred tier where regional variation is plausible. A conservative interpretation limited to the higher-confidence companies still identifies a substantial pattern warranting investigation.

Our prevalence estimate (77/123 companies) has a 95\% Wilson score confidence interval~\cite{wilson1927probable} of 53.4\%--71.1\%, reflecting sampling variability. For the instance count, inter-rater reliability analysis revealed a 1.1\% three-way disagreement rate (38 of 3,444 segments required consensus discussion), suggesting classification uncertainty of approximately $\pm$38 instances. Under conservative assumptions removing all disputed cases, the lower bound would be 244 instances across approximately 70 companies (56.9\%), suggesting that a majority of large companies in our sample may employ jurisdiction-siloed disclosure. This preliminary estimate supports our hypothesis that jurisdiction-siloed disclosure is a prevalent organizational pattern rather than an isolated practice, though independent replication is needed to validate these findings (see Section~\ref{sec:validation}).

A notable concentration of instances appears in platform companies with global infrastructure, raising the question of sensitivity to outliers. Roblox accounts for 41 of 282 instances (14.5\%), substantially higher than the next-largest contributor (Replit, 12 instances). To assess sensitivity to this outlier, we recalculated prevalence excluding Roblox: company-level prevalence remains 62.3\% (76 of 122 companies), with 241 instances across remaining companies. While instance count drops 14.5\%, company-level prevalence is robust to Roblox exclusion. This consistency suggests that jurisdiction-siloed disclosure is a systemic pattern rather than driven by a single company's practices.

Of the 123 companies in our sample, 105 (85.4\%) maintain jurisdiction-specific sections in their privacy policies. Among these 105 companies, 77 (73.3\%) exhibit jurisdiction-siloed disclosure: substantive practices disclosed only within regional sections. The remaining 28 companies with regional sections use them appropriately, containing only procedural information about exercising jurisdiction-specific rights while disclosing substantive practices universally. Notably, 18 companies (14.6\% of our sample) maintain no jurisdiction-specific sections at all; these include smaller organizations without multinational compliance infrastructure, companies operating primarily in single jurisdictions, and, in a few cases, organizations that appear to practice universal disclosure by default, making no structural distinction between users based on geography.

\begin{table}[t]
\caption{Jurisdiction-siloed disclosures by category}
\label{tab:siloed_categories}
\small
\begin{tabular}{@{}lrrr@{}}
\toprule
\textbf{Category} & \textbf{Regional} & \textbf{Intl.} & \textbf{Total} \\
\midrule
First-Party Collection & 51 & 22 & 73 \\
Sale/Sharing & 63 & 6 & 69 \\
Third-Party Sharing & 38 & 27 & 65 \\
Sensitive Data & 38 & 3 & 41 \\
Automated Decisions & 34 & 0 & 34 \\
\midrule
\textbf{Total} & \textbf{224} & \textbf{58} & \textbf{282} \\
\bottomrule
\end{tabular}

\vspace{0.5em}
\parbox{\linewidth}{\raggedright\footnotesize\textit{Note:} ``Regional'' = US-specific sections (California CCPA, Illinois BIPA); ``Intl.'' = non-US jurisdictional sections (EU/GDPR, UK). Instance counts reflect 1.1\% classification uncertainty ($\pm$38 instances).}
\end{table}

First-party data collection and data sales are the most commonly siloed categories, with collection practices and sales appearing in California sections as required by CCPA, often without equivalent specific disclosure elsewhere in the policy. Third-party sharing is notably prevalent in international sections, reflecting GDPR's disclosure requirements about data recipients. Automated decision-making disclosures appear only in domestic regional sections (primarily California), despite GDPR Article~22's requirements, suggesting companies may address automated decisions through other mechanisms for EU users or disclose such practices elsewhere in EU-specific documents.

\subsection{Case Studies}

To illustrate how jurisdiction-siloed disclosure operates in practice, we present five case studies drawn from our corpus. These cases demonstrate varying levels of verification: \textit{verified} cases involve direct evidence that disclosed practices apply globally (Clearview AI); \textit{strongly inferred} cases involve universality reasoned from documented global operations and technical infrastructure (Roblox, Northrop Grumman); \textit{moderately inferred} cases involve universality reasoned from policy language combined with known business models (Grindr, Thomson Reuters). We selected these five cases to represent each verification tier while spanning distinct industries (facial recognition, gaming, dating, defense contracting, data services) and different disclosure categories (biometric data, sensitive data, automated decisions, data sales). We emphasize that our contribution is identifying the \textit{disclosure pattern}---jurisdiction-specific sections that provide qualitatively more detailed, structured, and definitive disclosures than general sections covering the same practices---rather than proving specific data practices apply identically across jurisdictions. In several cases, general sections mention relevant practices in qualified or conditional language, but jurisdiction-specific sections provide structured confirmations, enumerated categories, and actionable rights absent from the general text. This asymmetry creates meaningful information gaps regardless of whether underlying practices are universal or reflect legitimate regional variation. Each case study follows a consistent structure: company context, general section language, jurisdiction-specific section language, and analysis of the resulting disclosure gap.

\subsubsection{Clearview AI: Biometric Terminology Confined to Illinois Section \emph{(Verified)}}

\emph{Company context.} Clearview AI operates a facial recognition database containing billions of images scraped from the public internet. It has been fined by data protection authorities in the Netherlands, France, Italy, Greece, Austria, and the United Kingdom, and its universal data collection is confirmed through extensive public reporting, Congressional testimony, and ongoing litigation.

\emph{General section language.} The general policy references ``information derived from the facial appearance of individuals in the photos'' and mentions ``face vector data'' in a retention section.

\emph{Jurisdiction-specific section language.} The term ``biometric''---with its associated legal significance and consumer recognition---appears exclusively in an Illinois-specific section referencing BIPA opt-out rights.

\emph{Analysis.} The general policy's circumlocutory language avoids the terminology that would signal biometric processing to a typical reader. Under the navigation patterns discussed in Section~\ref{sec:methodology}, a user in New York or Germany would encounter no use of the legally significant term ``biometric'' in the main policy sections, no biometric-specific rights, and no opt-out mechanism for facial recognition processing---only the oblique references noted above. This case is uniquely verified through external documentation confirming universal data collection.

\subsubsection{Grindr: Sensitive Data Collection Framed Conditionally Outside California \emph{(Inferred)}}

\emph{Company context.} Grindr is a dating app for LGBTQ+ users, handling inherently sensitive data categories including sexual orientation, precise geolocation, and health information.

\emph{General section language.} The general policy mentions sensitive data types using qualified, conditional language, noting these ``may be considered `sensitive' or `special category' Personal Information'' in ``some jurisdictions.''

\emph{Jurisdiction-specific section language.} The California section provides structured, definitive confirmation:

\begin{quote}
``In the past 12 months, we have collected the categories of personal information and sensitive personal information...''
\end{quote}

This California-specific table itemizes each sensitive data category (including sexual orientation and precise geolocation) with enumerated collection purposes.

\emph{Analysis.} We acknowledge that Grindr's conditional framing may be descriptively accurate for U.S.\ federal law, which lacks a comprehensive sensitive data classification. However, the general section's characterization of sensitivity as jurisdiction-dependent is difficult to reconcile with the broader legal landscape: sexual orientation data is classified as a special or sensitive category under the GDPR (Article~9), the CCPA/CPRA, Brazil's LGPD, South Africa's POPIA, and Council of Europe Convention~108+. Grindr's general-section language treats as a matter of jurisdictional perspective what these frameworks treat as an objective data classification. The structured California disclosure, which confirms sensitive data collection as fact, demonstrates that the company itself recognizes the applicability of this classification where law compels it. A non-California user navigating by section headers would encounter only the hedged framing, without the definitive confirmation or enumerated detail that the California section provides. This framing creates transparency asymmetry for users in the eighteen states with sensitive data categories and for users subject to international frameworks where sexual orientation is categorically protected. The question is whether users outside compulsory jurisdictions should receive equivalent substantive information about practices that affect them identically.

\subsubsection{Thomson Reuters: Data Sales Disclosed Definitively Only in California Section \emph{(Inferred)}}

\emph{Company context.} Thomson Reuters is a major data and information services provider whose business model involves extensive data aggregation and distribution.

\emph{General section language.} The general section acknowledges that providing services ``may constitute a `sale' of personal information'' under ``some local privacy laws.''

\emph{Jurisdiction-specific section language.} The California section provides explicit tables confirming that Thomson Reuters sells personal information across ten of twelve CCPA-defined categories, with granular opt-out rights.

\emph{Analysis.} An Ohio user reading only the general section would encounter equivocal language about potential data sales rather than the definitive categorical disclosure available to California residents. The general section's hedged ``may constitute a `sale'\,'' framing contrasts sharply with the California section's structured confirmation of sales across nearly all data categories. Both this case and Grindr demonstrate how jurisdiction-specific compliance creates substantive transparency gaps: general sections acknowledge practices in qualified terms while jurisdiction-specific sections provide the structured, actionable detail that constitutes meaningful transparency.

\subsubsection{Northrop Grumman: Automated Hiring Disclosure Confined to International Sections \emph{(Strongly Inferred)}}

\emph{Company context.} Northrop Grumman is a major defense contractor that uses AI-assisted tools in its recruitment process across jurisdictions.

\emph{General section language.} The U.S.\ applicant notice acknowledges AI use but states ``such technologies are not used to make automated decisions.''

\emph{Jurisdiction-specific section language.} The international notice (EU/UK and select other jurisdictions) reveals that ``AI may assist in making decisions during the recruitment process, such as shortlisting candidates or ranking applications.''

\emph{Analysis.} Non-EU job applicants are left without the right to reject, object to, or request explanations regarding automated hiring decisions, because the substantive disclosure of AI-assisted decision-making appears only in sections targeting jurisdictions that require it. The U.S.\ notice's assertion that AI is ``not used to make automated decisions'' contrasts with the international notice's acknowledgment that AI ``assist[s] in making decisions,'' creating a disclosure asymmetry for applicants subject to the same recruitment infrastructure.

\subsubsection{Roblox: Children Receive Less Transparency Than EU Residents \emph{(Strongly Inferred)}}

\emph{Company context.} Roblox is a gaming platform with over 70 million daily active users, predominantly children. It contains 41 jurisdiction-siloed instances---the highest in our corpus---spanning all four substantive categories: FIRST\_PARTY (14), THIRD\_PARTY (12), AUTOMATED\_DECISIONS (9), and SENSITIVE\_DATA (6). While Roblox's policy is more granular than comparable platforms (127 segments vs.\ 48--52 for Epic Games and Activision), excluding Roblox entirely changes company-level prevalence by only 0.3 percentage points, demonstrating the systemic pattern does not depend on this outlier.

\emph{General section language.} The general section contains a brief statement that the platform personalizes ads.

\emph{Jurisdiction-specific section language.} The EU/GDPR-specific section provides substantively more detailed disclosure:

\begin{quote}
``We process Personal Information to personalize the ads you see on Roblox. Specifically, we use information we collect and receive from partners to decide which ads to show you on Roblox.''
\end{quote}

The phrase ``and receive from partners'' appears only in the EEA version. The EEA section further discloses practices entirely absent from the general section: measuring and improving advertising effectiveness, and marketing via social media platforms to acquire new users.

\emph{Analysis.} A child user in the United States navigating by section headers would see a basic statement about ad personalization but would not encounter disclosures about partner data sourcing, cross-platform marketing, or advertising measurement---substantive details appearing only in EEA-specific sections. COPPA requires parental consent but does not mandate the detailed disclosure GDPR does, creating a transparency gap that jurisdiction-siloed organization exploits. This case also illustrates an inherent methodological limitation: two interpretations are possible. Either Roblox processes all users identically but discloses practices only where GDPR mandates transparency (a transparency failure), or Roblox implements genuinely different data handling under GDPR and appropriately discloses these differences only in EU sections (legitimate variation). Technical inspection or regulatory records would be needed to distinguish these scenarios.

\subsection{Industry Patterns}

\begin{table}[!htb]
\caption{Jurisdiction-siloed disclosure by industry}
\label{tab:industry_patterns}
\begin{tabular}{lccc}
\toprule
\textbf{Industry} & \textbf{Companies} & \textbf{\%} & \textbf{95\% CI} \\
\midrule
Big Tech & 6/7 & 86\% & 42--98\% \\
AI/ML & 7/9 & 78\% & 45--94\% \\
E-commerce & 7/9 & 78\% & 45--94\% \\
Financial Services & 9/12 & 75\% & 47--91\% \\
Surveillance/Defense & 7/10 & 70\% & 40--89\% \\
Data Brokers & 5/8 & 62\% & 30--87\% \\
Healthcare & 3/6 & 50\% & 19--81\% \\
\bottomrule
\end{tabular}

\vspace{0.5em}
\parbox{\linewidth}{\raggedright\footnotesize\textit{Note:} 95\% Wilson score confidence intervals reflect substantial uncertainty due to small industry subsamples.}
\end{table}

Exploratory analysis by industry reveals apparent variation in siloing rates, though small subsamples preclude confident conclusions. Table~\ref{tab:industry_patterns} reports rates ranging from 50\% (Healthcare, $n=6$) to 86\% (Big Tech, $n=7$), but 95\% confidence intervals span 40--60 percentage points for most industries. With the largest subsample containing only 12 companies (Financial Services), these patterns should be interpreted as preliminary observations suggesting hypotheses for future research rather than robust findings about industry-level behavior.

With that caveat, two patterns merit discussion. First, Big Tech's high company-level prevalence (86\%) may be an artifact of policy length rather than siloing intensity. Big Tech policies average 44.2 segments versus 27.2 for other companies, yet their siloing rate \emph{per segment} is actually lower (3.5\% vs.\ 8.3\%). Larger policies provide more opportunities for at least one siloed instance to occur; 89.4\% of all siloed instances in our corpus come from non-Big-Tech companies. Policy complexity alone does not drive siloing; if anything, more sophisticated legal operations produce better-organized policies with proportionally fewer siloed disclosures.

Second, Healthcare's lower observed rate (50\%) suggests a hypothesis worth testing with larger samples: sector-specific regulations like HIPAA, which mandate uniform disclosure regardless of patient jurisdiction, may constrain the organizational latitude that enables jurisdiction-siloed disclosure elsewhere. Alternatively, heightened public sensitivity to health privacy may motivate more careful policy organization. Distinguishing these explanations requires industry-specific samples large enough for meaningful statistical comparison.

\subsection{The Compounding Effect}

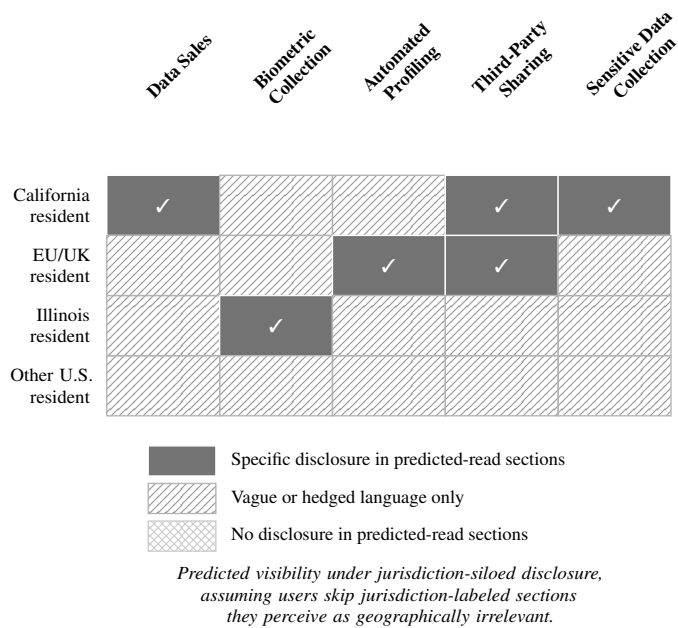
\begin{figure}[t]
\centering
\begin{tikzpicture}[
    every node/.style={font=\scriptsize},
]

\tikzset{
    specific/.style={fill=black!55, text=white, minimum width=1.5cm, minimum height=0.8cm, align=center, draw=white, line width=0.5pt},
    vague/.style={pattern=north east lines, pattern color=black!40, minimum width=1.5cm, minimum height=0.8cm, align=center, draw=black!30, line width=0.5pt},
    absent/.style={pattern=crosshatch, pattern color=black!20, minimum width=1.5cm, minimum height=0.8cm, align=center, draw=black!20, line width=0.5pt},
    header/.style={font=\scriptsize\bfseries, minimum width=1.5cm, minimum height=0.9cm, align=center, rotate=45, anchor=south west},
    rowlabel/.style={font=\scriptsize, minimum height=0.8cm, align=right, anchor=east, text width=2.2cm},
}

\node[header] at (0.75, 3.8) {Data Sales};
\node[header] at (2.25, 3.8) {Biometric\\Collection};
\node[header] at (3.75, 3.8) {Automated\\Profiling};
\node[header] at (5.25, 3.8) {Third-Party\\Sharing};
\node[header] at (6.75, 3.8) {Sensitive Data\\Collection};

\node[rowlabel] at (-0.1, 2.8) {California\\resident};
\node[rowlabel] at (-0.1, 2.0) {EU/UK\\resident};
\node[rowlabel] at (-0.1, 1.2) {Illinois\\resident};
\node[rowlabel] at (-0.1, 0.4) {Other U.S.\\resident};

\node[specific] at (0.75, 2.8) {\cmark};
\node[vague]    at (2.25, 2.8) {};
\node[vague]    at (3.75, 2.8) {};
\node[specific] at (5.25, 2.8) {\cmark};
\node[specific] at (6.75, 2.8) {\cmark};

\node[vague]    at (0.75, 2.0) {};
\node[vague]    at (2.25, 2.0) {};
\node[specific] at (3.75, 2.0) {\cmark};
\node[specific] at (5.25, 2.0) {\cmark};
\node[vague]    at (6.75, 2.0) {};

\node[vague]    at (0.75, 1.2) {};
\node[specific] at (2.25, 1.2) {\cmark};
\node[vague]    at (3.75, 1.2) {};
\node[vague]    at (5.25, 1.2) {};
\node[vague]    at (6.75, 1.2) {};

\node[vague]    at (0.75, 0.4) {};
\node[vague]    at (2.25, 0.4) {};
\node[vague]    at (3.75, 0.4) {};
\node[vague]    at (5.25, 0.4) {};
\node[vague]    at (6.75, 0.4) {};

\node[specific, minimum width=0.9cm, minimum height=0.4cm] (leg1) at (1.0, -0.6) {};
\node[font=\scriptsize, right=2pt of leg1] {Specific disclosure in predicted-read sections};
\node[vague, minimum width=0.9cm, minimum height=0.4cm] (leg2) at (1.0, -1.1) {};
\node[font=\scriptsize, right=2pt of leg2] {Vague or hedged language only};
\node[absent, minimum width=0.9cm, minimum height=0.4cm] (leg3) at (1.0, -1.6) {};
\node[font=\scriptsize, right=2pt of leg3] {No disclosure in predicted-read sections};

\node[font=\scriptsize\itshape, text width=7cm, align=center] at (3.75, -2.4) {Predicted visibility under jurisdiction-siloed disclosure,\\assuming users skip jurisdiction-labeled sections\\they perceive as geographically irrelevant.};

\end{tikzpicture}
\caption{Predicted information visibility by user jurisdiction under jurisdiction-siloed disclosure. Each cell represents the disclosure quality a user in a given jurisdiction would encounter for a given practice category. This is a theoretical illustration of predicted effects, not measured behavior; the behavioral assumption that users skip jurisdiction-labeled sections they perceive as geographically irrelevant, while grounded in information foraging theory (Section~\ref{sec:behavioral_validation}), remains unvalidated through direct observation of privacy policy navigation. Solid dark cells with checkmarks indicate specific, structured disclosure; diagonal hatching indicates vague or hedged language; crosshatch pattern indicates no disclosure. Grayscale shading and pattern fills provide redundant encoding for accessibility. Users outside regulated jurisdictions (bottom row) would encounter systematically less specific information despite being subject to the same data practices, if the predicted navigation patterns obtain. This illustrative pattern is derived from the disclosure asymmetries documented in our case studies (Section~5.2) and should not be read as representing any single company's policy.}
\label{fig:visibility_gradient}
\end{figure}

Jurisdiction-siloed disclosure is particularly problematic because it compounds existing information asymmetries. Users in jurisdictions with strong privacy laws benefit twice: they have legal rights \emph{and} receive more complete disclosure. Users in jurisdictions without such laws are doubly disadvantaged: they lack rights \emph{and} receive incomplete information about practices affecting them.

Empirical research establishes that users engage minimally with privacy policies and skip content they perceive as irrelevant. Obar and Oeldorf-Hirsch~\cite{obar2020biggest} found that 74\% of participants skipped privacy policies entirely when given the option to ``quick join.'' Among those who did access policies, average reading time was only 73 seconds, sufficient to read approximately 300 words at average reading speed, yet the policies in their study averaged nearly 8,000 words requiring 29--32 minutes for thorough reading. Users thus engage with only a small fraction of policy content. Critically, their follow-up research found users explicitly describe privacy policies as ``irrelevant'' to their needs~\cite{oeldorfhirsch2019overwhelming}, viewing them as overwhelming yet simultaneously irrelevant~\cite{obar2022older}.

We hypothesize that this perception of irrelevance, already a barrier for privacy policies generally, is amplified when section headings signal geographic inapplicability. A user already allocating only 73 seconds to a policy, and already viewing it as irrelevant, has strong reason to skip a section titled ``Your California Privacy Rights'' if they do not reside in California. The structural conditions established in the literature for content that users systematically overlook are satisfied: headings that signal irrelevance, content that requires affirmative action to reveal, and framing that discourages engagement (see Section~\ref{sec:behavioral_validation} for validation requirements).

This structural problem is further compounded by UI rendering choices. Our collection methodology required browser automation for 62\% of policies, many of which used collapsed accordion interfaces. Critically, sections titled ``California,'' ``For California Residents,'' or similar jurisdiction-specific headings are frequently rendered collapsed by default; a user actually reading the policy in their browser may not see the applicable terms unless they manually expand each section. A Texas user might reasonably skip a collapsed ``California Privacy Rights'' section assuming it contains only California-specific procedures, never suspecting it contains the only specific disclosure that the company sells their data.

This creates structural conditions that function as barriers to notice: the disclosure exists but (1) appears in a section where users skip with higher probability than universal sections, (2) is collapsed by default requiring affirmative user action to reveal, and (3) contains information framed as jurisdiction-specific when the underlying practice is universal. If this pattern produces the predicted effects on user awareness, it would create a surveillance inequality where the most vulnerable users---those without legal protection---are also the least informed about practices they cannot contest.

\section{Implications}

\subsection{For Regulators}

Our analysis suggests three areas for regulatory clarification.

First, the FTC's ``clear and conspicuous'' standard, requiring disclosures that are ``difficult to miss,'' focuses on user perception rather than mere textual presence~\cite{ftc2021negative}. The FTC's 2022 staff report on dark patterns explicitly taxonomizes ``design elements that hide or delay disclosure of material information'' as deceptive practices~\cite{ftc2022darkpatterns}, a framing directly applicable to jurisdiction-siloed disclosure. A disclosure placed where users predictably skip based on geographic framing may not satisfy this standard. Regulators could clarify whether organizational placement affects conspicuousness assessments. Recent enforcement actions confirm regulatory attention to structural asymmetry in privacy interfaces. In January 2022, France's CNIL fined Google \texteuro150 million for cookie consent interfaces requiring multiple clicks to refuse cookies while offering single-click acceptance; the CNIL found that this structural asymmetry ``affects the freedom of consent.'' Similar reasoning produced \texteuro60 million fines against Microsoft (December 2022) and \texteuro5 million against TikTok (January 2023) for comparable interface asymmetries. In the United States, the FTC's 2023 complaint against Amazon alleged that Prime cancellation required four pages, six clicks, and fifteen options compared to two clicks for enrollment, a case resulting in a \$2.5 billion settlement in September 2025. While these cases address consent interfaces and cancellation flows rather than privacy policy disclosure structure, they establish the principle that structural design choices affecting how easily users can access or act on privacy information constitute potential violations of transparency requirements.

Second, GDPR's WP260 guidance that layered notices should prominently present ``processing which could surprise'' suggests that substantive disclosures should not appear exclusively in regional sections. European regulators could clarify whether jurisdiction-siloed disclosure of surprising practices conflicts with transparency principles.

Third, regulators could require that any substantive practice disclosed in one jurisdiction be disclosed universally. This would not expand disclosure \emph{content}; it would require existing disclosures to appear where all affected users can find them, eliminating compliance arbitrage.

We document a structural pattern that undermines transparency; the question of whether it violates existing law involves statutory interpretation beyond our expertise. Standards bodies could specify that substantive disclosures must appear in universal sections, enabling consistent evaluation across companies.

\subsection{For Researchers}

Section~\ref{sec:validation} details the validation studies needed: methodological replication, independent human annotation, and behavioral testing. The research community can also contribute to developing structural transparency standards and apply our released tools to larger corpora. The \emph{privacy.txt} specification~\cite{colwell2024privacytxt}, an IETF Internet-Draft proposing machine-readable privacy disclosures, would enable automated detection of jurisdiction-siloed patterns at scale.

\section{Call for Community Validation}
\label{sec:validation}

This preprint is released to invite systematic validation of the jurisdiction-siloed disclosure phenomenon. We identify three priority areas for independent research.

\subsection{Methodological Replication}

As discussed in Section~\ref{sec:circularity}, our classification rules were developed on the same corpus used for analysis. Independent researchers should:

\begin{itemize}
    \item Apply our released classification prompt to new samples of privacy policies not included in our corpus
    \item Compare prevalence estimates across fresh samples to assess generalizability
    \item Test whether the phenomenon appears at similar rates in different company populations (e.g., small businesses, international companies, specific industries)
\end{itemize}

If our methodology is not subject to the circularity bias discussed in Section~\ref{sec:circularity}, we predict that jurisdiction-siloed disclosure will appear in fresh samples at rates statistically indistinguishable from our preliminary estimate (see Section~\ref{sec:prevalence} for confidence interval). Failure to replicate would suggest our classification rules overfit to our specific corpus, confirming the circularity concern.

\subsection{Independent Human Annotation}

As noted in Section~\ref{sec:opp115_comparison}, our methodology provides consistency but not the independent validity that human annotation would provide. Independent validation must:

\begin{itemize}
    \item Recruit multiple human annotators (ideally with legal or privacy expertise) to classify a sample of our segments
    \item Measure inter-annotator reliability among human coders
    \item Compare human consensus labels to our LLM-derived labels
    \item Identify systematic disagreements that reveal classification ambiguities
\end{itemize}

We release segment-level data including the original text, heading paths, and our assigned labels to facilitate this validation. Cohen's Kappa between human annotators and our methodology would provide validity evidence that our OPP-115 comparison (Section~\ref{sec:classification}) cannot, since OPP-115 lacks the post-2018 categories central to our taxonomy.

\subsection{Behavioral Validation}
\label{sec:behavioral_validation}

Our argument that jurisdiction-siloed disclosure undermines notice rests on an informed inference from established behavioral research: information foraging theory predicts that users skip content whose headings signal irrelevance~\cite{pirolli1999information}, empirical studies document selective heading-based navigation in privacy policies~\cite{vu2007users}, and 74\% of users skip policies entirely when possible~\cite{obar2020biggest}. Additional evidence supports the relevance-based skipping mechanism: eye-tracking research documents that readers scan headings while skipping body text when headings signal irrelevance~\cite{buscher2009surfing}; topic headings facilitate processing of relevant content while creating gating decisions for irrelevant sections~\cite{hyona2004topic}; and bounded rationality models predict that users satisfice rather than maximize information acquisition under cognitive load~\cite{acquisti2015privacy}. Geographic labels in section headings (e.g., ``California Residents,'' ``European Users'') function as explicit negative relevance signals; information foraging theory predicts these trigger skipping behavior among users outside the named jurisdictions. These converging lines of evidence strongly predict that users outside a named jurisdiction will skip sections labeled for that jurisdiction. While no study has specifically measured skipping rates for jurisdiction-labeled sections, the inference is grounded in multiple independent research traditions rather than speculation. Direct measurement would strengthen this foundation and quantify the effect size.

The critical question for validation is not merely whether users skip jurisdiction-labeled sections, but whether the effect size is large enough to materially reduce notice effectiveness. Even a modest differential attention pattern, if consistent across the user population, could substantially reduce the proportion of affected users who receive specific disclosure of practices affecting them. Eye-tracking research on privacy policies~\cite{steinfeld2016agree} demonstrates that users who open policies skim rather than read, focusing on headings and first sentences of paragraphs to assess relevance. This heading-driven navigation pattern suggests that geographic framing in section titles may function as a strong relevance signal, though the magnitude of this effect for jurisdiction-specific sections remains unquantified.

Estimating the required effect size: if 74\% of users skip policies entirely~\cite{obar2020biggest} and the remaining 26\% who engage employ heading-guided skimming as Steinfeld~\cite{steinfeld2016agree} documents, even a modest 50\% differential skipping rate for jurisdiction-labeled sections among active readers would mean approximately 13\% of all users (half of the 26\% who read) miss jurisdiction-siloed disclosures they would otherwise encounter. For substantive practices affecting millions of users, this represents a material transparency gap. Future validation studies should be powered to detect differential skipping rates of at least this magnitude.

Validation should:

\begin{itemize}
    \item Conduct eye-tracking studies comparing attention to jurisdiction-specific versus universal sections
    \item Measure whether users in non-triggering jurisdictions (e.g., Texas residents encountering ``California Privacy Rights'') skip such sections at measurably higher rates
    \item Test whether users who read entire policies including regional sections exhibit better comprehension of disclosed practices than users who skip regional content
\end{itemize}

Negative findings---evidence that users do not systematically skip jurisdiction-labeled sections---would undermine the transparency concern motivating this paper, even if the organizational pattern exists.

\subsection{What Would Disconfirm Our Findings?}

We specify conditions under which we would consider our hypothesis disconfirmed:

\begin{enumerate}
    \item Replication failure: Independent researchers applying our methodology to fresh samples find jurisdiction-siloed disclosure in fewer than 40\% of major companies (outside our confidence interval).
    \item Classification invalidity: Human annotators substantially disagree with our labels ($\kappa < 0.4$), suggesting our classification rules do not capture expert judgment about the substantive/procedural distinction.
    \item Behavioral disconfirmation: User studies find no significant difference in attention to jurisdiction-labeled versus universal sections, undermining the notice concern.
    \item Universality refutation: Evidence that practices disclosed in regional sections genuinely vary by jurisdiction (e.g., technical documentation showing companies maintain jurisdiction-specific data processing pipelines) for the majority of our identified instances.
\end{enumerate}

We invite researchers to pursue these disconfirmation conditions. A hypothesis worth proposing is a hypothesis worth testing.

\section{Discussion}

\subsection{Intent vs. Effect}

We do not claim that companies deliberately adopt this organizational pattern with deceptive intent. The structural pattern likely emerges from legitimate compliance motivations: legal teams structure policies around regulatory requirements, placing CCPA content in California sections and GDPR content in EU sections.

But intent does not determine effect. Whatever the motivation, users outside regulated jurisdictions may receive systematically incomplete information about practices affecting them, to the extent they navigate by section headers as information foraging theory predicts. Universal substantive disclosure addresses this potential effect regardless of whether it was intended.

Our classification reflects the judgment of a single domain expert encoded in a shared prompt and refined through iterative boundary resolutions during corpus annotation. Following standard practice in content analysis for exploratory research~\cite{oconnor2020intercoder}, we acknowledge that single-coder annotation prioritizes consistent application of expert judgment but cannot demonstrate the convergent validity that multiple independent annotators would provide. While three-model agreement (Fleiss' $\kappa$ = 0.858) demonstrates reproducibility, this consistency reflects prompt clarity rather than validated correctness (three LLMs applying the same instructions consistently is expected if the instructions are unambiguous). OPP-115 validation (Cohen's $\kappa$ = 0.629) confirms alignment with human judgment for pre-2016 categories, but the post-2018 categories central to our findings await independent human annotation. We frame these classifications as hypothesis-generating rather than definitive, inviting the validation described in Section~\ref{sec:validation}.

\subsection{Template Propagation}

The prevalence of jurisdiction-siloed disclosure likely reflects systematic propagation through compliance infrastructure rather than independent organizational choices by individual companies. Privacy compliance platforms dominate enterprise policy authoring: OneTrust alone holds approximately 30\% market share and serves 75\% of the Fortune 100~\cite{onetrust2024marketshare}; the broader privacy management software market exceeded \$5 billion in 2025~\cite{fortunebi2025privacy}. These platforms provide template-based policy generators that structure jurisdiction-specific sections as discrete policy regions. OneTrust offers privacy notice templates that ``pre-populate notice sections with real-time OneTrust data'' and provide ``automated notification guidance per jurisdiction''~\cite{onetrust2024templates}. TrustArc provides over 800 operational templates with ``jurisdiction-specific information and considerations''~\cite{trustarc2024templates}.

Academic research supports the template propagation hypothesis. An analysis of privacy policy similarity found that Jaccard Similarity Indexes greater than 0.7 ``are likely due to identical privacy policies only differing by organization name'' and that ``similarity between groups of privacy policies is likely the result of privacy policy generators or privacy policy templates''~\cite{brashear2023similarity}. The structural similarity evidence, combined with documented template features organizing jurisdiction-specific sections as discrete policy regions, is consistent with the hypothesis that the organizational pattern we observe propagates through compliance infrastructure rather than emerging independently at each company. These tools map regulatory requirements to policy architecture, creating sections titled ``Your California Privacy Rights'' or ``Information for EU Residents'' that contain substantive disclosures mandated by CCPA or GDPR but not required elsewhere. While we cannot definitively confirm the causal mechanism linking template availability to the specific organizational patterns we observe, several lines of circumstantial evidence converge: (1) market dominance of template-based compliance platforms, (2) documented template features that structure jurisdiction-specific sections as discrete regions, (3) academic findings of high policy similarity consistent with template use~\cite{brashear2023similarity}, and (4) structural consistency across unrelated companies in our sample. This evidence supports templated origins as a plausible explanation, though direct validation would require comparing policies from companies with documented template use against those known to draft policies independently.

This propagation mechanism has important implications. First, it suggests that jurisdiction-siloed disclosure is not primarily a strategic choice by individual companies but an emergent property of compliance infrastructure. Addressing the pattern may require engaging with template vendors and industry standards bodies, not only individual companies. Second, the same infrastructure that enables jurisdiction-siloed disclosure could readily support universal substantive disclosure; templates could be redesigned to place substantive content universally while confining regional sections to procedural information. Third, the template-based origin suggests that regulatory guidance clarifying disclosure architecture could propagate through the compliance ecosystem as efficiently as the current problematic pattern emerged. A single FTC interpretation clarifying that ``clear and conspicuous'' requires universal placement of substantive disclosures could cascade through vendor templates to thousands of policies.

\subsection{The Limits of Notice}

Our findings compound existing critiques~\cite{cranor2012necessary} that notice is necessary but insufficient for meaningful privacy protection. Cranor argued that even well-designed standardized notice mechanisms cannot overcome fundamental limitations of the notice-and-choice paradigm; our structural findings suggest an additional failure mode: organizational choices that undermine notice effectiveness independent of content quality or standardization. Even if policies were shorter, clearer, and written at accessible reading levels, organizational choices could still undermine transparency. A perfectly readable disclosure offers limited value if it appears only in a section that many users would likely skip based on heading cues.

This suggests that structural requirements, not only content requirements, are necessary for meaningful notice. Regulating \emph{what} companies must disclose is insufficient if companies can choose \emph{where} to disclose it.

\subsection{Toward Structural Transparency}

We advocate for a shift from content-focused to structure-focused transparency requirements. Current regulations specify what must be disclosed; future regulations should specify how disclosures must be organized to ensure visibility.

Universal substantive disclosure offers one structural standard. Others might include requirements for summary sections, standardized organization, or layered disclosure formats. The common thread is attention to document architecture as a transparency mechanism.

Prior work demonstrates that structured formats can meaningfully improve privacy comprehension. Kelley et al.'s ``privacy nutrition label'' enables users to find information faster and more accurately than natural language policies, with standardized grids allowing comparison across services~\cite{kelley2009nutrition}. Machine-readable formats offer complementary benefits: the proposed \texttt{privacy.txt} standard~\cite{colwell2024privacytxt} would enable automated auditing of disclosure practices, though the current draft specifies structural requirements without addressing substantive disclosure placement. A natural extension would require that \texttt{privacy.txt} declarations explicitly indicate whether disclosed practices apply universally or are limited to specific jurisdictions, operationalizing universal substantive disclosure at the technical layer.

Our contribution operates at this architectural level: identifying organizational choices that create conditions where information discovery is impeded, as grounded in the behavioral research discussed in Section~\ref{sec:methodology}. The structural pattern warrants transparency scrutiny regardless of the precise magnitude of its behavioral effects; even if users skip jurisdiction-labeled sections at lower rates than theory predicts, organizational choices that place substantive disclosures exclusively in geographically-framed sections represent a departure from the transparency principles motivating disclosure requirements.

\section{Conclusion}

Privacy policies are supposed to provide notice. Our audit of 123 major companies suggests a structural pattern that may undermine this function: substantive disclosures about data sales, biometric collection, and automated profiling appearing in specific, structured form only within jurisdiction-specific sections, while general sections provide at most qualified or conditional language for the same practices.

Our preliminary analysis suggests nearly two-thirds of sampled companies exhibit jurisdiction-siloed disclosure, with 62.1\% of identified instances falling into Strongly or Moderately Inferred tiers where legitimate regional variation is implausible. If the behavioral assumptions grounding this work are validated (Section~\ref{sec:behavioral_validation}), users in unregulated jurisdictions would be unlikely to encounter specific information about substantive practices affecting them through typical navigation patterns, receiving notice that is systematically less detailed based on geography rather than actual data practices.

As discussed in Section~\ref{sec:methodology}, our analysis cannot definitively distinguish between universal practices disclosed only where legally required (a transparency failure) and genuinely jurisdiction-specific practices appropriately disclosed in regional sections (legitimate variation). While evidence favors the transparency-failure interpretation for most instances (Section~\ref{sec:methodology}), this work identifies patterns warranting investigation, not definitive proof of transparency violations.

As detailed in Section~\ref{sec:opp115_comparison}, our OPP-115 validation applies only to pre-2016 categories; the post-2018 categories central to our findings lack independent human validation. A conservative estimate limited to validated categories encompasses 138 instances (49\% of identified cases), still supporting the prevalence hypothesis but with more modest magnitude.

We propose \emph{universal substantive disclosure}: substantive practices affecting all users should be disclosed universally, unless a company explicitly states that a practice applies only to certain jurisdictions. Regional sections should contain procedural information about exercising jurisdiction-specific rights and, when genuinely applicable, disclosure of practices that occur only in those jurisdictions. This standard does not require uniform legal protection; it requires that truly universal practices be disclosed universally and that practice limitations be made explicit rather than implicit through selective disclosure.

The notice-and-choice framework assumes notice is provided. Our findings suggest that when disclosures exist but are placed in locations users are likely to avoid based on established navigation patterns, notice may fail even for users who read carefully. Structural transparency---attention to how policies are organized, not only what they contain---is necessary for meaningful privacy disclosure.

We release our methodology and dataset to enable independent validation, replication, and ongoing audits. The question is not only what companies disclose, but whether anyone can find it---and whether the pattern we identify withstands independent scrutiny.

\section*{Data Availability}

The annotated dataset of 3,651 privacy policy segments from 123 companies (corpus identifier: \texttt{OPPT-T1\_\allowbreak C1.0\_\allowbreak Section\_\allowbreak Jan2026}) is available at:\footnote{Inter-rater reliability analysis in Section~\ref{sec:methodology} was performed on 3,444 segments; 207 segments were excluded from reliability calculations due to incomplete three-model classifications or quality filtering, though all 3,651 segments are included in the released dataset with appropriate metadata flags.}
\begin{center}
\url{https://huggingface.co/datasets/OpenPrivacyPolicyTaxonomy/oppt-privacy-policies}
\end{center}
The dataset includes segment text, hierarchical heading paths, primary and secondary category labels, consensus type (unanimous, majority, or expert-resolved), and annotation methodology metadata. The taxonomy and corpus are versioned independently; the taxonomy is \texttt{OPPT~v1.0} and future corpus releases may update the policy collection or annotation models without changing the taxonomy. The dataset is released under CC BY-NC 4.0; commercial licensing inquiries may be directed to the corresponding author.

The OPPT v1.0 taxonomy specification, annotation methodology, machine-readable attribute schema, and example analysis code are available at:
\begin{center}
\url{https://github.com/Open-Privacy-Policy-Taxonomy/oppt}
\end{center}


\bibliography{references}

\appendix

\section{Companies with Jurisdiction-Siloed Disclosure}

\begin{table}[ht]
\caption{Top 20 companies by siloed disclosure instances}
\label{tab:companies}
\small
\begin{tabular}{lcc>{\raggedright\arraybackslash}p{3.8cm}}
\toprule
\textbf{Company} & \textbf{Inst.} & \textbf{Verif.} & \textbf{Categories Siloed} \\
\midrule
Roblox & 41 & \cmark & FIRST\_PARTY, THIRD\_PARTY, AUTOMATED\_DECISIONS, SENSITIVE\_DATA \\
Replit & 12 & --- & AUTOMATED\_DECISIONS, SALE\_SHARING \\
PayPal & 11 & H & FIRST\_PARTY, AUTOMATED\_DECISIONS, THIRD\_PARTY, SENSITIVE\_DATA \\
Verizon & 10 & H & SALE\_SHARING, SENSITIVE\_DATA, AUTOMATED\_DECISIONS \\
Hilton & 9 & --- & FIRST\_PARTY, THIRD\_PARTY \\
Epic Games & 8 & H & THIRD\_PARTY, FIRST\_PARTY, SALE\_SHARING \\
Grindr & 7 & --- & SALE\_SHARING, FIRST\_PARTY, AUTOMATED\_DECISIONS \\
Redfin & 7 & --- & SENSITIVE\_DATA, SALE\_SHARING, FIRST\_PARTY \\
BetterHelp & 6 & --- & THIRD\_PARTY, SENSITIVE\_DATA, FIRST\_PARTY \\
Booking & 6 & --- & THIRD\_PARTY, SALE\_SHARING, SENSITIVE\_DATA \\
Duolingo & 6 & --- & FIRST\_PARTY, THIRD\_PARTY \\
Microsoft & 6 & H & FIRST\_PARTY, THIRD\_PARTY, SENSITIVE\_DATA \\
Stripe & 6 & H & AUTOMATED\_DECISIONS, SENSITIVE\_DATA \\
Zillow & 6 & --- & SALE\_SHARING, FIRST\_PARTY, SENSITIVE\_DATA \\
Alibaba & 5 & --- & FIRST\_PARTY, THIRD\_PARTY, AUTOMATED\_DECISIONS \\
DraftKings & 5 & --- & SALE\_SHARING, SENSITIVE\_DATA, THIRD\_PARTY \\
L3Harris & 5 & --- & THIRD\_PARTY, SALE\_SHARING, FIRST\_PARTY \\
Bumble & 4 & --- & SENSITIVE\_DATA, FIRST\_PARTY \\
Disney & 4 & H & FIRST\_PARTY, THIRD\_PARTY, SALE\_SHARING \\
Eyematch AI & 4 & --- & SALE\_SHARING, FIRST\_PARTY, SENSITIVE\_DATA \\
\bottomrule
\end{tabular}

\vspace{0.5em}
\parbox{\linewidth}{\raggedright\footnotesize\textit{Note:} Verification status: \cmark = verified through external documentation (investor disclosures, regulatory filings, litigation); H = strongly inferred (global platform infrastructure); --- = moderately or weakly inferred from policy language and business model. See Table~\ref{tab:confidence_tiers} for inference tier definitions.}
\end{table}

\emph{Full list of 77 companies available in supplementary materials.}

\FloatBarrier
\section{Classification Categories}

We classified policy segments using a 14-category taxonomy derived from OPP-115~\cite{wilson2016creation} and extended to address post-2018 regulatory developments:

\begin{description}
\item[FIRST\_PARTY] Data the company collects directly and how it is used.
\item[THIRD\_PARTY] Data shared with or received from external parties.
\item[USER\_CHOICE] Opt-outs, preferences, and consent management mechanisms.
\item[USER\_ACCESS] Rights to access, correct, delete, or port personal data.
\item[RETENTION] How long data is kept and deletion schedules.
\item[SECURITY] Encryption, safeguards, and breach notification procedures.
\item[POLICY\_CHANGE] How users are notified of policy updates.
\item[TRACKING] Cookies, beacons, analytics, targeted advertising, and GPC/DNT signals.
\item[INTL\_SPECIFIC] Children's privacy (COPPA) and international data transfers.
\item[OTHER] Introduction, definitions, contact information, and boilerplate.
\item[REGIONAL] Jurisdiction-specific sections (California Privacy Rights, EU/UK Users, Illinois Residents) containing procedures for exercising legal rights.
\item[SALE\_SHARING] Disclosures about whether the company sells personal data or shares it for cross-context behavioral advertising.
\item[AUTOMATED\_DECISIONS] Profiling, algorithmic decisions, automated eligibility determinations, or AI-driven processing.
\item[SENSITIVE\_DATA] Practices involving biometric data, health information, genetic data, precise geolocation, or other sensitive categories.
\end{description}

\medskip

For identifying jurisdiction-siloed disclosure, we focus on six categories. A disclosure is classified as ``siloed'' when substantive content (FIRST\_PARTY, THIRD\_PARTY, SALE\_SHARING, SENSITIVE\_DATA, or AUTOMATED\_DECISIONS) appears within a REGIONAL section without equivalent disclosure in a universal section of the policy.

\subsection{Substantive vs. Procedural Coding Rules}

Applying the substantive/procedural distinction defined in Section~\ref{sec:distinction}, we provide operational examples for annotation:

\emph{Substantive examples}: ``We sell your personal information to third parties''; ``We collect biometric identifiers including facial geometry''; ``We use automated decision-making to determine eligibility''; ``We share your data with advertising partners.''

\emph{Procedural examples}: ``To opt out, click the link below''; ``You may request access to your data by emailing \url{privacy@company.com}''; ``California residents can submit a deletion request through our portal''; ``To file a complaint, contact the Irish Data Protection Commission.''

\emph{Borderline cases}: When procedural rights imply substantive practices (e.g., ``Right to opt out of sale'' implies sales occur), we code the segment as substantive if the underlying practice lacks universal disclosure; the user learns about the practice only because they encountered the procedural section.

\subsection{Category Boundary Distinctions}

Prompt refinement during corpus annotation produced eight documented boundary distinctions that resolve recurring ambiguities. These rules ensure consistent classification across the three-model ensemble.

\begin{enumerate}
    \item THIRD\_PARTY vs.\ SALE\_SHARING: Segments containing ``sale,'' ``sell,'' or ``Do Not Sell'' language are classified as SALE\_SHARING; operational sharing with service providers, partners, or affiliates without sale terminology is THIRD\_PARTY.

    \item USER\_CHOICE vs.\ USER\_ACCESS: Preferences, opt-outs, and consent mechanisms are USER\_CHOICE; data subject rights (access, correction, deletion, portability) are USER\_ACCESS.

    \item REGIONAL vs.\ substantive categories: Segments in jurisdiction-specific sections describing \emph{procedures} for exercising rights are REGIONAL; segments describing \emph{practices}, even within regional sections, are classified by their substantive content.

    \item INTL\_SPECIFIC vs.\ REGIONAL: Children's privacy (COPPA compliance) and international data transfers are INTL\_SPECIFIC; jurisdiction-specific rights procedures (California, EU, state-specific sections) are REGIONAL.

    \item TRACKING vs.\ FIRST\_PARTY: When tracking technologies (cookies, pixels, fingerprinting) are the focus, classify as TRACKING; when general data collection is described with incidental mention of tracking, classify as FIRST\_PARTY.

    \item SENSITIVE\_DATA vs.\ FIRST\_PARTY: When special categories of data (biometric, health, genetic, precise geolocation) are the explicit focus, classify as SENSITIVE\_DATA; when sensitive data types appear incidentally in a broader collection statement, classify as FIRST\_PARTY.

    \item SECURITY vs.\ OTHER: Company security practices (encryption, access controls, breach procedures) are SECURITY; user-facing security advice (``keep your password safe'') is OTHER.

    \item AUTOMATED\_DECISIONS vs.\ OTHER: Substantive descriptions of algorithmic processing, profiling, or AI-driven decision-making are AUTOMATED\_DECISIONS; AI-related platitudes (``committed to responsible AI'') without substantive disclosure are OTHER.
\end{enumerate}

\subsection{Classification Prompt}

The complete classification prompt used for three-model annotation is available in the dataset repository.\footnote{\url{https://huggingface.co/datasets/OpenPrivacyPolicyTaxonomy/oppt-privacy-policies}} The prompt includes category definitions for all 14 OPPT categories, the eight boundary rules enumerated above, required attribute annotations per OPP-115 schema, and an anti-hiding instruction specifying that content in regional sections describing practices (not procedures) is classified by substance, not as REGIONAL.

The prompt underwent iterative refinement during annotation of the 123-policy corpus. Initial versions achieved approximately 78\% inter-model agreement on complex policies; the final prompt achieved 93\%+ agreement. For reproducibility, we recommend researchers use the exact prompt version archived with the dataset rather than adapting these rules independently, as subtle phrasing differences may affect classification consistency.

\section{Methodology Details}

\subsection{Collection Protocol}

Policies were collected January 2026 via three methods:
\begin{itemize}
    \item Direct HTTP request (35.8\%)
    \item Browser automation with JavaScript rendering (61.8\%)
    \item Internet Archive retrieval for blocked sites (2.4\%)
\end{itemize}

\subsection{Three-Model Classification}

Classification prompts were iteratively refined by the author, a certified privacy professional (CIPP/US, AIGP) with five years of enterprise data governance experience, during annotation of the full 123-policy corpus. This refinement resolved approximately 280 classification disputes during iterative prompt development and produced eight documented boundary distinctions. Three frontier LLMs from different providers---Claude Haiku 4.5 (Anthropic), GPT-5.2 (OpenAI), and Gemini-3-flash (Google)---independently classified each segment using the shared prompt.

\begin{table}[ht]
\caption{Inter-model agreement metrics}
\label{tab:agreement}
\begin{tabular}{lc}
\toprule
\textbf{Metric} & \textbf{Value} \\
\midrule
Unanimous agreement (3/3) & 78.3\% \\
Majority agreement (2/3) & 20.5\% \\
Model consensus (usable) & 98.7\% \\
Claude$\leftrightarrow$GPT & 83.8\% \\
Claude$\leftrightarrow$Gemini & 85.5\% \\
GPT$\leftrightarrow$Gemini & 86.1\% \\
\bottomrule
\end{tabular}
\end{table}

Final labels were determined by majority vote; the 1.3\% of segments without majority agreement were resolved by the author. Using three models from different providers offers some protection against idiosyncratic model behavior, though we acknowledge these models likely share substantial training data and may exhibit similar blind spots for legal and regulatory text. Agreement among them could reflect shared training artifacts rather than classification validity. The consistency metrics demonstrate that the classification task produces reproducible results across different LLM architectures, but this reproducibility should not be conflated with validation against human ground truth.

\subsection{Siloed Disclosure Identification}

For each company, we:
\begin{enumerate}
    \item Identified all REGIONAL segments
    \item Extracted secondary category labels within REGIONAL segments
    \item Checked whether equivalent substantive disclosure appeared in non-REGIONAL sections
    \item Flagged instances where substantive content appeared \emph{only} in REGIONAL sections
\end{enumerate}

This conservative methodology captures only clear cases of jurisdiction-siloed disclosure.

\end{document}